\documentclass[reprint,superscriptaddress,prl,twocolumn,amsmath,amssymb,aps,longbibliography,hidelinks, colorlinks=true,citecolor=blue,linkcolor=blue,urlcolor=blue]{revtex4-2}

\usepackage{graphicx}
\usepackage{soul}
\usepackage{dcolumn}
\usepackage{bm}
\usepackage{hyperref}
\usepackage{xr-hyper}
\usepackage{xcolor}
\usepackage[normalem]{ulem}

\usepackage{mathtools}
\newcommand\underrel[3][]{\mathrel{\mathop{#3}\limits_{%
			\ifx c#1\relax\mathclap{#2}\else#2\fi}}}

\newcommand{\numcomponent}{n}

\newcommand{\ddint}[1]{\int d^d #1}

\newcommand{\nablasquared}{\nabla^2}
\newcommand{\hc}[1]{#1^{\dagger}}
\newcommand{\cc}[1]{\mkern 1.5mu\overline{\mkern-1.5mu#1\mkern-1.5mu}\mkern 1.5mu}
\newcommand{\Zset}{\mathbb{Z}}

\makeatletter
\newcommand{\symgroup}[1]{U(\@ifempty{#1}{n}{#1})}
\makeatother 

\newcommand{\Rset}{\mathbb{R}}

\newcommand{\gpvec}[1]{\mathbf{#1}}

\newcommand{\zerovec}{\gpvec{0}}
\newcommand{\xvec}{\gpvec{x}}

\newcommand{\kvec}{\gpvec{k}}

\newcommand{\phivec}{\bm{\phi}}

\newcommand{\psitilde}{\tilde{\psi}}

\newcommand{\chihat}{\hat{\chi}}

\newcommand{\Chat}{\hat{C}}

\newcommand{\psivec}{\bm{\psi}}
\newcommand{\psicomplex}{\cc{\psi}}
\newcommand{\psiveccomplex}{\hc{\psivec}}

\newcommand{\psitildecomplex}{\cc{\psitilde}}

\newcommand{\xivec}{\bm{\xi}}

\newcommand{\Exp}[1]{\operatorname{exp}\left(#1\right)}

\newcommand{\Erefs}[1]{\mbox{Eqs.~(\ref{eq:#1})}}

\newcommand{\kvecsquared}{\kvec^2}

\newcommand{\Hamiltonian}[1][]{\mathcal{H}_{#1}}

\newcommand{\phiveci}{\bm{\phi}_i}

\newcommand{\xiveccomplex}{\hc{\xivec}}
\newcommand{\xiveccomp}[1][]{\xi^{#1}}
\newcommand{\xiveccompcomplex}[1][]{\cc{\xi}^{#1}}

\newcommand{\psiveccomp}[1][]{\psi^{#1}}

\newcommand{\av}[1]{\langle #1 \rangle}

\newcommand{\condensation}{\rho_0}
\newcommand{\frequencylc}{\Omega}

\makeatletter
\newcommand{\Qtensor}[1]{Q^{\@ifempty{#1}{abcd}{#1}}}
\newcommand{\kroneckerdelta}[1]{\delta^{\@ifempty{#1}{ab}{#1}}}
\makeatother

\newcommand{\forpidtwo}{(\forpi)^{d/2}}

\newcommand{\forpi}{4 \pi}

\makeatletter
\newcommand{\invprop}[2]{\Gamma^{\@ifempty{#1}{ab}{#1}}_{\psitildecomplex \psi}(\@ifempty{#2}{\kvec, \omega}{#2})}
\newcommand{\invpropcomplex}[2]{\Gamma^{\@ifempty{#1}{ab}{#1}}_{\psitilde \psicomplex}(\@ifempty{#2}{\kvec, \omega}{#2})}
\newcommand{\vertexnoise}[2]{\Gamma^{\@ifempty{#1}{ab}{#1}}_{\psitildecomplex \psitilde}\@ifempty{#2}{}{(#2)}}
\newcommand{\vertexnonlin}[2]{\Gamma^{\@ifempty{#1}{abcd}{#1}}_{\psitildecomplex \psicomplex \psi \psi}\@ifempty{#2}{}{(#2)}}
\newcommand{\vertexnonlincomplex}[2]{\Gamma^{\@ifempty{#1}{abcd}{#1}}_{\psitilde \psi \psicomplex \psicomplex}\@ifempty{#2}{}{(#2)}}

\newcommand{\invpropren}[2]{\Gamma^{\@ifempty{#1}{ab}{#1}}_{R,\psitildecomplex \psi}(\@ifempty{#2}{\kvec, \omega}{#2})}
\newcommand{\invpropcomplexren}[2]{\Gamma^{\@ifempty{#1}{ab}{#1}}_{R, \psitilde \psicomplex}(\@ifempty{#2}{\kvec, \omega}{#2})}
\newcommand{\vertexnoiseren}[2]{\Gamma^{\@ifempty{#1}{ab}{#1}}_{R, \psitildecomplex \psitilde}\@ifempty{#2}{}{(#2)}}
\newcommand{\vertexnonlinren}[2]{\Gamma^{\@ifempty{#1}{abcd}{#1}}_{R, \psitildecomplex \psicomplex \psi \psi}\@ifempty{#2}{}{(#2)}}
\newcommand{\vertexnonlincomplexren}[2]{\Gamma^{\@ifempty{#1}{abcd}{#1}}_{R, \psitilde \psi \psicomplex \psicomplex}\@ifempty{#2}{}{(#2)}}

\makeatother

\newcommand{\diff}{D}
\newcommand{\diffp}{D'}
\newcommand{\mass}{r}
\newcommand{\massp}{r'}
\newcommand{\taup}{\tau'}
\newcommand{\noise}{\Gamma}
\newcommand{\nonlincoupling}{u}
\newcommand{\nonlincouplingp}{u'}

\newcommand{\ratiodiffR}{w_{D,R}}
\newcommand{\rationonlinR}{w_{u,R}}

\newcommand{\utilde}{\tilde{u}}

\newcommand{\utildestar}{\tilde{u}^*_{R}}
\newcommand{\wDstar}{\ratiodiffR^{*}}
\newcommand{\wustar}{\rationonlinR^{*}}

\makeatletter
\newcommand{\deltak}[1]{\Delta\left(\@ifempty{#1}{\kvec}{#1}\right)}
\newcommand{\deltakcomplex}[1]{\cc{\Delta}\left(\@ifempty{#1}{\kvec}{#1}\right)}
\makeatother

\newcommand{\imag}{\mathring{\imath}}

\begin{document}
\hypersetup{
    colorlinks=true,
    citecolor=blue,
    linkcolor=blue,
    urlcolor=blue
}

\title{Non-Reciprocal yet Equilibrium Critical Dynamics}
	
\author{Emir Sezik}%
	\email{emir.sezik19@imperial.ac.uk}
	\affiliation{Department of Mathematics, Imperial College London, South Kensington, London SW7 2AZ, United Kingdom}


\date{\today}

\begin{abstract}
Non-reciprocal interactions find broad applicability in non-equilibrium and living systems. Their canonical implementation involves asymmetric couplings between two entities, which generally induce spatio-temporal patterns and time-dependent steady states that break time-translational invariance, representing a clear deviation from equilibrium physics. Although their phenomenology is well understood, whether non-reciprocal interactions induce new universality classes, and if so, under what conditions, remains an open question. In the present work, we perform a field-theoretic renormalization group (RG) analysis of the dynamics of two non-reciprocally coupled $n$-vector order parameters possessing a $U(n)$ symmetry, generalizing previous results to order parameters with multiple components, a feature that has been shown to generate novel non-equilibrium critical behavior in certain non-reciprocal systems. To lowest order in $\epsilon = 4-d$, we find that the non-reciprocal coupling is RG-irrelevant, and the critical behavior is governed by the equilibrium fixed point of the Model A universality class of Hohenberg and Halperin with $2n$ vector components, even though the transition is into a non-equilibrium state. 
Our results demonstrate that non-reciprocity alone may not be sufficient to induce novel universality classes.
\end{abstract}

\maketitle

The breaking of action-reaction symmetry, or non-reciprocity, is ubiquitous in non-equilibrium and living systems, giving rise to a variety of collective phenomena. These include time-dependent steady states and spatio-temporal patterns \cite{you2020nonreciprocity, fruchart_non-reciprocal_2021, shankar_topological_2022, marchetti_nr_pattern, avni_dynamical_2024, avni_non-reciprocal_2024, lorenzana2024nonreciprocal, johnsrud_phase_2025, saha_scalar_2020, phase_coexistence_quorum_duan_2025, dinelli2023nonreciprocity, cocconi_active_bound_states_2023, symmetry_topology_thermodynamics_marchetti}, topological defect dynamics \cite{bandini_xy_2024, loos_long-range_2023, rouzaire_non-reciprocal_2024, dopierala_inescapable_2025,ordering_and_Defect_ramaswamy,liu_dynamic_2025}, synchronization \cite{risler_universal_2004, risler_universal_2005, osat_non-reciprocal_2023, nr_synchronisation_matteo, kreienkamp_synchronization_2025}, and the formation of time crystals \cite{daviet_kardar-parisi-zhang_2024, hanai_nonreciprocal_2024, daviet_nonequilibrium_2024, zelle_universal_2024}. Non-reciprocal interactions have also proven essential in quantitatively describing social dynamics \cite{reichenbach_mobility_2007, vicsekCollectiveMotion2012, bainDynamicResponseHydrodynamics2019}, living matter \cite{chen2025chirality, theveneauChaseandrunAdjacentCell2013,  paoluzziInformationMotilityExchange2020} and active solids \cite{scheibner_odd_2020, odd_elasticity_fruchart, fruchartOddViscosityOdd2023}, establishing them as a central concept in active matter and non-equilibrium physics in general.

Broadly, non-reciprocity refers to a scenario in which two entities interact asymmetrically, violating either Newton's Third Law or detailed balance. Due to this broad definition, many active matter systems can be labeled as non-reciprocal, as they generically violate detailed balance. However, most common implementation of non-reciprocity is to couple two fields in an antagonistic way \cite{you2020nonreciprocity, fruchart_non-reciprocal_2021, shankar_topological_2022, johnsrud_phase_2025, saha_scalar_2020, young_nonequilibrium_2020, young_nonequilibrium_2024, pisegna_emergent_2024}. The resulting behavior depends sensitively on the precise way non-reciprocity is implemented. If the fields are coupled non-reciprocally at least at the linear level, the system settles into a dynamical steady state and hosts spatio-temporal patterns \cite{you2020nonreciprocity, fruchart_non-reciprocal_2021, shankar_topological_2022, johnsrud_phase_2025, saha_scalar_2020,symmetry_topology_thermodynamics_marchetti, pisegna_emergent_2024}; whereas if the non-reciprocal interaction is only biquadratic in the fields, the system demonstrates regular condensation behavior at late times with no spatio-temporal patterns \cite{young_nonequilibrium_2020, young_nonequilibrium_2024, sezik2026criticaldynamicsnonreciprocallycoupled}. The ensuing ordering dynamics has also been shown to be case-dependent. For the former implementation, it was found that the transition into these time-dependent states is \emph{effectively thermal} in non-reciprocally coupled spin models \cite{lorenzana2025nonreciprocity, risler_universal_2004, risler_universal_2005,tauber_perturbative_2014, avni_dynamical_2024, avni_non-reciprocal_2024} and the universality class was found to fall within the Model A class of Halperin and Hohenberg's classification \cite{hohenberg_theory_1977}. This result is perhaps unsurprising given that the restoration of detailed balance at macroscopic scales has been demonstrated to be possible despite microscopic non-reciprocity in certain cases \cite{akritidis2026fateisinguniversalityclass, bassler_critical_1994, chen2025numerical, scandolo2026nonequilibriumcouplingdiffusingdensity}. Strikingly, biquadratic couplings, which exhibit exclusively condensation behavior, have been shown to induce a novel non-equilibrium universality class \cite{young_nonequilibrium_2020, young_nonequilibrium_2024}. At first glance, it might seem unintuitive that different microscopic implementations of non-reciprocity lead to different critical behavior, apparently at odds with universality. However, this is not the case as different types of interactions endow the system with different symmetries, which in turn determine their respective universality classes. For example, in Refs.~\cite{risler_universal_2004, risler_universal_2005,tauber_perturbative_2014}, where the fields are coupled non-reciprocally at the linear level, the model is symmetric under the group $U(1)$. By contrast, in Ref.~\cite{young_nonequilibrium_2020}, where the non-reciprocity enters only nonlinearly, the symmetry group of the model is $\Zset_2 \times \Zset_2$, a far more restrictive symmetry.

The role of symmetry in determining the universality class naturally raises the question of what happens when the symmetry group is enlarged — for instance, when the order parameter has multiple components. Thus far, the numerical and analytical studies of the relevance of non-reciprocity to the ordering dynamics have been exclusively focused on two single-component fields \cite{risler_universal_2004, risler_universal_2005, tauber_perturbative_2014, lorenzana2025nonreciprocity, avni_non-reciprocal_2024, avni_dynamical_2024, young_nonequilibrium_2020, guislain_collective_2024, sieberer_nonequilibrium_2014, sieberer_prl}, corresponding to the critical dynamics of non-reciprocally coupled Ising-like models, with Refs.~\cite{young_nonequilibrium_2024, sezik2026criticaldynamicsnonreciprocallycoupled} as the only exceptions. Importantly, Refs.~\cite{young_nonequilibrium_2024, sezik2026criticaldynamicsnonreciprocallycoupled} showed that the number of components of the fields plays a crucial role in determining the universality class of the ordering dynamics, where, for a sufficiently large number of vector components, non-reciprocity becomes irrelevant at large scales.

In this Letter, we consider the ordering dynamics of a class of non-reciprocally coupled vector fields, generalizing the discussion in Refs.~\cite{risler_universal_2004, risler_universal_2005, tauber_perturbative_2014} to $n$-component vector fields. The fields individually undergo Model A dynamics of Halperin and Hohenberg's classification \cite{hohenberg_theory_1977} but are then coupled in a non-reciprocal manner. We choose an implementation where the non-reciprocal couplings induce dynamical steady states, in contrast to Refs.~\cite{young_nonequilibrium_2020, young_nonequilibrium_2024, sezik2026criticaldynamicsnonreciprocallycoupled}. We first show that the effective theory describing the transition is a multi-component noisy complex Ginzburg-Landau equation \cite{garcia-morales_complex_2012, aranson_world_2002} in which the non-reciprocity is encoded in the imaginary couplings. By mapping the dynamical model onto a field theory using established techniques \cite{statistical_dynamics_martin_1973, janssen_lagrangean_1976,  bausch_renormalized_1976, janssen_renormalized_1977, janssen_renormalized_1992, tauber_critical_2014}, we perform the field-theoretic renormalization group (RG) procedure \cite{cardy1996scaling, zinn-justin_quantum_2002, bellac_quantum_1992, amit_field_2005,tauber_critical_2014} to lowest order in $\epsilon = d_c -d$, where $d_c$ is the upper critical dimension, to characterize its critical dynamics. In contrast to Ref.~\cite{young_nonequilibrium_2024}, we find that the non-reciprocal couplings are RG-irrelevant regardless of the number of field components. The non-reciprocity gives rise only to a subleading drive exponent (defined below) controlling the resonant frequency of the response function, a feature also observed in  \cite{daviet_nonequilibrium_2024, tauber_perturbative_2014}. 
Our results indicate that even though the transition is into an oscillatory state, the ordering dynamics of non-reciprocally coupled models with multiple degrees of freedom, such as XY or Heisenberg models, is effectively equilibrium, with the universality class falling under the Hohenberg and Halperin classification \cite{hohenberg_theory_1977}. Our results, combined with those of Refs. \cite{young_nonequilibrium_2024, tauber_perturbative_2014, lorenzana2025nonreciprocity, risler_universal_2004, risler_universal_2005}, show that non-reciprocity may not be sufficient to generate novel non-equilibrium universality classes. 

\textit{Model ---} We consider the ordering dynamics of two isotropic vector order parameters, $\phiveci \in \Rset^{\numcomponent}$ for $i \in\{1,2\}$, coupled in a non-reciprocal manner that induces ``run-and-chase" dynamics, in which one field chases the other while the other flees \cite{fruchart_non-reciprocal_2021}. The model is defined by the following Langevin equation
\begin{equation}\label{eq:modelv1main}
    \dot{\phivec}_i = -\frac{\delta \Hamiltonian}{\delta \phivec_i} - \sum_{j = 1}^{2}\epsilon_{ij}\frac{\delta \Hamiltonian'}{\delta \phivec_j} +  \sqrt{2\noise} \xivec_i
\end{equation}
where $\epsilon_{ij}$ is the Levi-Civita tensor satisfying $\epsilon_{11} = \epsilon_{22} = 0$ and $\epsilon_{12} = -\epsilon_{21} = 1$, $\xivec_i$ are 
independently and identically distributed Gaussian noises with zero mean and unit variance satisfying $\langle \xi_{i}^{a}(\xvec,t) \xi^{b}_j(\xvec',t')\rangle = \kroneckerdelta{} \delta_{ij}\delta(\xvec - \xvec') \delta(t-t')$ with the subscript $i$ denoting the species and $a,b \in\{1,2,\dots ,\numcomponent\}$ the vector component of the noise. Here,  $\Hamiltonian$ governs the relaxational (Model A) dynamics of each field, while $\Hamiltonian'$ encodes the inter-species non-reciprocal interactions, enforced by the Levi-Civita tensor $\epsilon_{ij}$. Their explicit forms are
\begin{subequations} \label{eq:Hamiltonian}
\begin{align} 
    \Hamiltonian &= \ddint{x}\left[\sum_{i = 1}^{2}\left\{ \frac{D}{2}(\nabla\phivec_i)^2 + \frac{r}{2}\phivec_i^2\right\} + \frac{u}{4!}(\phivec_1^2 + \phivec_2^2)^2 \right]\\
    \Hamiltonian' &= \ddint{x}\left[\sum_{i = 1}^{2}\left\{ \frac{D'}{2}(\nabla\phivec_i)^2 + \frac{r'}{2}\phivec_i^2\right\} + \frac{u'}{4!}(\phivec_1^2 + \phivec_2^2)^2 \right]
\end{align}
\end{subequations}
The parameters of the model have the following physical roles. The usual parameters $r,D$ and $u > 0$ represent, respectively, the field stiffness, the diffusion coefficient, and the quartic self-interaction, and $\noise$ sets the noise strength. The non-reciprocal parameters $r', D'$ and $u'$ control, respectively, the linear non-reciprocal coupling, the antagonistic diffusion, and the nonlinear non-reciprocal interaction between the two fields. For stability, we require $u'D'>0$ \cite{SuppMatt}; when this condition is violated, the antagonistic diffusion and nonlinear coupling have competing effects, leading to a finite-wavelength instability in the ordered phase \cite{garcia-morales_complex_2012}. The $u'(\phivec_1^2 + \phivec_2^2)^2$ term in $\Hamiltonian'$ and the reciprocal nonlinear interaction $u\phivec_1^2\phivec_2^2$ between the fields in $\Hamiltonian$ are required for RG consistency: even if absent from the bare theory, such terms are generated under renormalization and must therefore be retained from the outset. For $n=2$ and $n=3$, Eqs.~\eqref{eq:modelv1main} may describe, for example, the dynamics of two non-reciprocally coupled XY and Heisenberg models, respectively. Both fields fluctuate and interact in an identical manner, and are driven by noise with the same statistics, ensuring that any deviation from equilibrium universality, should there be one, arises solely from the non-reciprocal couplings.

Even though we have chosen the bare model to have no time or length-scale separation between the two fields, such a separation could in principle be generated under RG flow where a single coupling, say $u'$, might acquire different corrections, spoiling the form of Eqs.~\eqref{eq:modelv1main}. However, this does not happen as the form of \Erefs{modelv1main} is protected by the symmetry of the model. By requiring the fields to obey the same dynamics, quantified by the Hamiltonians $\Hamiltonian[]$ and $\Hamiltonian[]'$, we have endowed the system with the symmetry group $U(\numcomponent)$, the unitary group. To make this explicit, we define the two complex vector order parameters 
\begin{equation}
    \psivec \equiv (\phivec_1 + \imag \phivec_2)/\sqrt{2}   \qquad\psiveccomplex \equiv (\phivec^{T}_1 - \imag \phivec^{T}_2)/\sqrt{2}
\end{equation}
where $\phi_i^{T}$ denotes the transpose of $\phivec_i$, making $\psivec^\dagger$ a row vector. We then rewrite Eqs.~\eqref{eq:modelv1main} in terms of these new \emph{linearly independent} fields
\begin{subequations} \label{eq:modelv2main}
    \begin{align}
        \dot{\psivec} &= - \left\{ r+\imag r' -(\diff + \imag\diffp)\nablasquared + \frac{u + \imag u'}{3} \psiveccomplex\cdot \nonumber 
        \psivec\right\} \psivec  \\
        &+ \sqrt{2\Gamma}\xivec \label{eq:modelv2amain}\\
        \dot{\psiveccomplex} &= -\left\{ r-\imag r' -(\diff - \imag \diffp)\nablasquared + \frac{u - \imag u'}{3} \psiveccomplex\cdot \psivec\right\} \psiveccomplex \nonumber \\
        &+\sqrt{2\Gamma}\xiveccomplex \label{eq:modelv2bmain}
    \end{align}
\end{subequations}
where $\xivec = (\xivec_1 + \imag \xivec_2)/\sqrt{2}$ is a complex Gaussian noise, satisfying $\av{\xiveccompcomplex[a](\xvec,t) \xiveccomp[b](\xvec',t')} =\kroneckerdelta{} \delta(\xvec - \xvec') \delta(t-t')$ and $\av{\xiveccomp[a](\xvec,t) \xiveccomp[b](\xvec',t')} = 0$. The two equations are related by complex conjugation, a structure that is preserved under RG flow. Within this parametrization, the $U(n)$ invariance is explicit, guaranteeing that the RG procedure does not generate terms that break the structure of Eqs.~\eqref{eq:modelv1main}.

Notably, these equations take the form of a noisy multi-component complex Ginzburg-Landau equation \cite{garcia-morales_complex_2012, aranson_world_2002}, a central equation in the description of pattern formation, nonlinear waves, and other phenomena such as superconductivity and Bose-Einstein condensation \cite{cross_pattern_1993, kuramoto1984chemical}. Setting $n=1$ recovers the model studied in \cite{tauber_perturbative_2014, risler_universal_2004, risler_universal_2005}, corresponding to the semi-classical dynamics of a driven-dissipative Bose-Einstein condensate \cite{sieberer_nonequilibrium_2014, sieberer_prl}. Setting $r'/r = \diffp/\diff = u'/u = \text{const.}$ recovers the model of \cite{de_dominicis_field-theoretic_1975}, in which the non-reciprocal forces become reversible and the system relaxes to a thermal steady state with Hamiltonian $\Hamiltonian$ \cite{tauber_perturbative_2014}.

Finally, it is instructive to compare the symmetry of our model with that of Refs.~\cite{young_nonequilibrium_2020, young_nonequilibrium_2024, sezik2026criticaldynamicsnonreciprocallycoupled}, where non-reciprocity is implemented only at the nonlinear level. In that case, the two fields remain fundamentally distinct, as neither can be continuously rotated into the other while leaving the dynamics invariant \footnote{Specifically, in our case $\phivec_1 \to \cos(\theta) \phivec_1 - \sin(\theta) \phivec_2$ and $\phivec_2 \to \cos(\theta) \phivec_2 + \sin(\theta) \phivec_1$ for constant $\theta$ is a valid transformation whereas it is not in the model considered in Refs.~\cite{young_nonequilibrium_2024}}. This is further captured by their symmetry group, $O(n) \times O(n)$, of dimension $n(n-1)$. By contrast, the identical functional form of $\Hamiltonian$ and $\Hamiltonian'$ in our model allows such rotations, enlarging the symmetry to $U(n)$ of dimension $n^2$. This enhanced symmetry is the key structural difference between our model and those of Refs.~\cite{young_nonequilibrium_2020, young_nonequilibrium_2024}, and we expect it to play an important role in determining the universality class of the critical dynamics.

\textit{Mean-Field Transition ---} Starting from Eqs.~\eqref{eq:modelv2main}, we can characterize the mean-field transition by considering a spatially uniform solution that solves them in the limit of vanishing noise. In \cite{SuppMatt}, choosing the ansatz $\psivec(\xvec,t) = \sqrt{\condensation}e^{-\imag \frequencylc t} \psivec_0$, where $\psivec_0$ is a constant complex vector satisfying $\psiveccomplex_0 \cdot \psivec_0 = 1$, we show that the linearly \emph{stable} ordered phases are controlled by the sign of $r$ and are given by 
\begin{equation} \label{eq:meanfieldtransition}
    (\condensation, \frequencylc) = \begin{cases}
        (0,0), ~\text{for} ~  r >0 \\
        \left( -\frac{3r}{u}, r'- r \frac{u'}{u}\right), ~\text{for} ~ r<0 
    \end{cases}
\end{equation}
where the stability of these solutions is guaranteed by the condition $u'D'>0$. The development of a non-zero order parameter amplitude $\rho_0 > 0$ for $r<0$ in a specific direction corresponds to the well-known spontaneous breaking of rotational symmetry in equilibrium \cite{zinn-justin_quantum_2002, amit_field_2005}, breaking $U(n)$ down to $U(n-1)$. In the present setting, however, the ordered phase also sustains global uniform oscillations at a finite frequency $\Omega$, further breaking time-translational symmetry, the latter being impossible in equilibrium. The onset of the ordered state is controlled by $r$ while $r'$ determines the oscillation frequency of the resulting steady state. For certain parameters, $r$ and $r'$, the oscillation frequency can be tuned to zero and the system only develops a non-zero order parameter amplitude. 
The breaking of time-translational invariance does not introduce additional Goldstone modes beyond those expected from the $U(n) \to U(n-1)$ breaking as a time translation $t \to t+ T$ acts on the order parameter as $\psivec \to \Exp{-\imag \Omega T}\psivec$, which is identical to the action of the broken $U(1)$ generator \cite{watanabe_redundancies_2013, nielsen_how_1976}.

\textit{RG analysis ---} We study the system near criticality by simultaneously tuning $r\to 0$ and $r'\to 0$. While tuning $r'\to 0$ is not strictly necessary as it can always be gauged away through a suitable field transformation, it represents a convenient choice in the RG analysis \cite{SuppMatt} and does not affect the end result. We characterize the critical dynamics using renormalized field theory with minimal subtraction and dimensional regularization \cite{tauber_critical_2014, zinn-justin_quantum_2002, bellac_quantum_1992, amit_field_2005}. Standard power-counting \cite{tauber_critical_2014} identifies the upper critical dimension as $d_c = 4$ \cite{SuppMatt}, and we therefore work in $d = 4-\epsilon$ dimensions throughout.

The ordering dynamics is characterized by the response function $\chi(\xvec,t) \kroneckerdelta{} = \delta\langle \psiveccomp[a](\xvec,t)\rangle/\delta h^b(\zerovec,0)|_{\bm{h}(\xvec,t) = 0}$, where $\bm{h}(\xvec,t)$ is a source field, and the correlation function $C(\xvec,t) \equiv  \langle \psiveccomplex(\xvec,t) \cdot \psivec(\zerovec,0)\rangle/n$. For Gaussian fluctuations, they take the form \cite{SuppMatt}:
\begin{subequations} \label{eq:baretheory}
    \begin{align}
        \chi(\kvec,\omega) &= \frac{1}{-\imag(\omega - r' -\diffp \kvecsquared) + \mass + \diff\kvecsquared} \label{eq:bare_response}\\
        C(\kvec,\omega) &= \frac{2 \noise}{\left( \omega - \massp - \diffp \kvecsquared\right)^2+ \left( \mass + \diff \kvecsquared\right)^2} \label{eq:bare_corr}
    \end{align}
\end{subequations}
where $\massp + \diffp\kvecsquared$ controls the resonant frequency of both functions and $\mass + \diff \kvecsquared$ controls their static behavior. Indeed, integrating the correlation function $C(\kvec,\omega)$ over $\omega$, we obtain the standard Ornstein-Zernike form for the equal time correlation function \cite{tauber_critical_2014}. When the system approaches criticality, Eqs.~\eqref{eq:baretheory} no longer hold as the system displays non-Gaussian fluctuations. Instead, the functions can be written in the following scaling form \cite{SuppMatt}:
\begin{subequations} \label{eq:scalingformsmain}
    \begin{align}
    &\chi(\kvec,\omega) = \frac{1}{|\kvec|^{2-\eta}} \chihat\left(\frac{\omega}{|\kvec|^{z}}, a |\kvec|^{\eta-\eta_c},\tau |\kvec|^{-1/\nu},  \taup |\kvec|^{-1/\nu'}\right), ~ \\
   & C(\kvec,\omega) = \frac{1}{|\kvec|^{2 + z-\eta'}}\times \nonumber\\
   & \qquad \Chat\left(\frac{\omega}{|\kvec|^{z}}, a|\kvec|^{\eta'-\eta_c},\tau |\kvec|^{-1/\nu}, \taup |\kvec|^{-1/\nu'}\right)
    \end{align}
\end{subequations}
where $\chihat$ and $\Chat$ are dimensionless functions, $\tau = r-r_c$ and $\tau'= r'-r'_c$ are the deviations from the critical point and $a$ is a non-universal constant. 

The exponents $z, \nu, \nu',\eta, \eta'$ and $\eta_c$ characterize the critical dynamics: $z$ is the dynamical exponent, $\nu$ and $\nu'$ control the divergence of the correlation lengths associated with $r$ and $r'$ respectively, and $\eta, \eta'$ and $\eta_c$ are the anomalous dimensions of the correlation and response functions. In non-equilibrium systems, the system is typically characterized by two independent scaling exponents, $\eta$ and $\eta'$, as there is no fluctuation-dissipation to enforce their equality \cite{tauber_critical_2014}. The non-reciprocity additionally induces a novel exponent $\eta_c$, which controls the resonant frequency of the response function \cite{tauber_perturbative_2014, daviet_nonequilibrium_2024}. At mean-field level ($d>d_c = 4$), one has $z = 2$, $\nu = \nu'= 1/2$ and $\eta = \eta' = \eta_c = 0$. For $d<d_c = 4$, the nonlinearities in Eqs.~\eqref{eq:Hamiltonian} become relevant and induce nontrivial scaling exponents.

\begin{figure}
    \centering
    \includegraphics[width=0.9\linewidth]{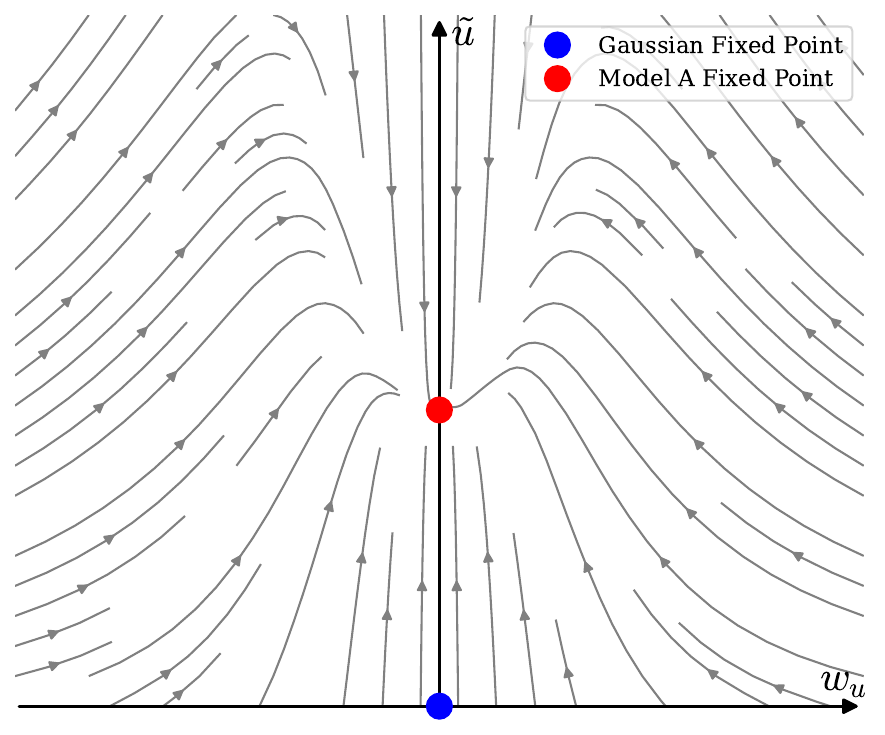}
    \caption{\textit{RG flows for $\utilde$ and $w_{u}$ at $w_D = 0$ ---} There are two fixed points present in the system: the trivial Gaussian fixed point and the Model A fixed point. For $d <4$, the system flows to the stable Model A fixed point, where the ordering dynamics is effectively thermal. The RG flows are odd in $w_u$ as the physics is overall invariant under swapping the signs of all the non-reciprocal couplings, which simply determine the direction of the ``run-and-chase" dynamics.}
    \label{fig:fig2}
\end{figure}

The scaling exponents can be obtained through a perturbative RG procedure \cite{cardy1996scaling, bellac_quantum_1992, amit_field_2005, zinn-justin_quantum_2002, tauber_critical_2014}. In \cite{SuppMatt}, we show that the perturbative expansion, and consequently the fixed points of the RG flow, are controlled by the following dimensionless parameters:
\begin{equation} \label{eq:expansion_parameters}
    \utilde \equiv \frac{\nonlincoupling \noise}{\forpidtwo \diff^2},~~w_u \equiv \frac{\nonlincouplingp}{\nonlincoupling}, ~~w_D = \frac{\diffp}{\diff}
\end{equation}
where $w_u$ measures the relative strength of $u'$ compared to $u$ and $w_D$ the relative strength of antagonistic diffusion to regular diffusion. In \cite{SuppMatt}, we calculate the flow functions associated with the parameters in Eq.~\eqref{eq:expansion_parameters} to lowest non-trivial order. Solving for the fixed points of the flow functions, we find that the stable fixed point below the upper critical dimension is given by the equilibrium Model A fixed point \cite{SuppMatt}:
\begin{equation} \label{eq:stablefixedpoint}
    \utildestar = \frac{3\epsilon}{2(n+4)}, ~~\wustar = \wDstar = 0.
\end{equation}
where the subscript $R$ corresponds to the renormalized counterparts of the couplings. The overall flow diagram for $w_D = 0$ is illustrated in Fig.~\ref{fig:fig2}.

Strikingly, the non-reciprocal couplings vanish at the stable fixed point, and the critical dynamics is governed entirely by the Model~A universality class. In contrast to Refs.~\cite{young_nonequilibrium_2024, daviet_nonequilibrium_2024}, where the number of field components alters the universality class, we find that the non-reciprocity becomes irrelevant regardless of the number of field components $n$. Furthermore, since $w^{*}_u = w^{*}_D = 0$, the complex field equations reduce to real equations with an enlarged and emergent $O(2n)$ symmetry at the critical point. The effective equilibrium dynamics at the fixed point also enforces $\eta = \eta'$, consistent with the restoration of detailed balance.

The main results of our RG analysis are the following scaling exponents at the fixed point, Eq.~\eqref{eq:stablefixedpoint}, which characterize the critical dynamics 
\begin{subequations}
\begin{align}
    z &= 2 + \frac{n+1}{4(n+4)^2}\left(6 \ln\left(\frac{4}{3}\right) - 1 \right) \epsilon^2 + \mathcal{O}(\epsilon^3),\\
    \eta &= \eta' = \frac{n+1}{4(n+4)^2}\epsilon^2+ \mathcal{O}(\epsilon^3),\\
    \eta_c &= -\frac{n+1}{4(n+4)^2}\left(4 \ln\left(\frac{4}{3}\right) - 1 \right)\epsilon^2 + \mathcal{O}(\epsilon^3),~ \\
    \nu &= \frac{1}{2} + \frac{n+1}{4(n+4)}\epsilon + \mathcal{O}(\epsilon^2), \\
    \nu' &= \frac{1}{2} + \mathcal{O}(\epsilon^2)
\end{align}
\end{subequations}
The critical exponents coincide with those of Model A \cite{tauber_critical_2014} for a $2n-$component real order parameter, consistent with the emergent $O(2n)$ at the fixed point. Notably, the correlation length exponent $\nu'$ associated with the non-reciprocal parameter $r'$ retains its mean-field value, in contrast to $\nu$ which is corrected at order $\epsilon$. We also observe that $\eta -\eta_c > 0$ for all $n$, so that at large length scales the equilibrium contribution $|\kvec|^{2-\eta}$ dominates over the non-reciprocal correction $|\kvec|^{2-\eta_c}$ in the response and correlation functions. Physically, this confirms that the effects of non-reciprocity are subleading compared to the standard Model~A behavior at the critical point.

In \cite{SuppMatt}, we also analyzed the model in the $\numcomponent \to \infty$ limit, where the scaling exponents can be computed exactly, finding that they remain equal to their equilibrium counterparts \cite{zinn-justin_quantum_2002, mosheQuantumFieldTheory2003a}, $z = 2$, $\eta = \eta' = \eta_c = 0$, and $\nu = 1/(d-2)$, and that $\nu'$ again acquires no correction.

\textit{Conclusions \& outlook---} In this Letter, we have studied the critical dynamics of non-reciprocally coupled vector order parameters with $n$ components, generalizing previous single-component analyses \cite{risler_universal_2004, risler_universal_2005, tauber_perturbative_2014} to order parameters with multiple degrees of freedom. We have shown that near the critical point, the effective dynamics is described by a noisy multi-component complex Ginzburg-Landau equation, with the non-reciprocity encoded in the imaginary coupling constants. In the ordered phase, the system displays homogeneous oscillations, simultaneously breaking $U(n) \to U(n-1)$ and time-translational symmetry. Using a perturbative field-theoretic RG scheme, we have shown that the non-reciprocal couplings are RG-irrelevant regardless of the number of field components, and that the critical behavior is governed by the Model A universality class of Hohenberg and Halperin \cite{hohenberg_theory_1977}. Strikingly, the non-reciprocal couplings vanish during the ordering dynamics, where the symmetry is effectively enlarged from $U(n) \to O(2n)$.
Our results, combined with previous works in the literature \cite{tauber_perturbative_2014, lorenzana2025nonreciprocity, risler_universal_2004, risler_universal_2005}, highlight the robustness of the Hohenberg and Halperin classification to non-reciprocity.

Future work will address the stability of this model to perturbations that break the $U(n)$ symmetry by assigning distinct coupling constants to the reciprocal and non-reciprocal parts of the Hamiltonian. This induces spatial and temporal scale separation between the fields, explicitly breaking the $U(n)$ symmetry down to $O(n)$; the effects of this reduction on the RG flow and resulting macroscopic behavior are not immediately known. A more comprehensive investigation of the types of non-reciprocal interactions that give rise to novel universality classes is also needed to classify the effects of non-reciprocity on the critical dynamics of a system.

\begin{acknowledgments}
\textit{Acknowledgements~---}~ES would like to thank Giulia Pisegna for interesting discussions and Gunnar Pruessner for invaluable support and interesting discussions. ES is supported by the Roth Scholarship at Imperial College London.
\end{acknowledgments}

\bibliography{NRwithU_n_symmetry}
	
\end{document}


\preprint{APS/123-QED}

\title{Supplementary Material for ``Non-Reciprocal yet Equilibrium Critical Dynamics"}

  \author{Emir Sezik}%
	\email{emir.sezik19@imperial.ac.uk}
	\affiliation{Department of Mathematics, Imperial College London, South Kensington, London SW7 2AZ, United Kingdom}


\date{\today}


\maketitle

\hrule
\vspace{-3\baselineskip} 
\tableofcontents
\vspace{1em}
\hrule

\section{Mean-Field Analysis} \label{sec:MeanField}
In this section, we perform a mean-field analysis to identify the different phases of the model. We start with the Langevin Eqs.~(4) of the main text 
\begin{subequations} \label{eq:modelv2supp} 
\begin{align}
    \dot{\psivec}(\xvec,t) &= - \left\{ r+\imag r' -(\diff + \imag\diffp)\nablasquared + \frac{u + \imag u'}{3} \psiveccomplex(\xvec,t)\cdot  
        \psivec(\xvec,t)\right\} \psivec(\xvec,t) + \sqrt{2\Gamma}\xivec(\xvec,t) \label{eq:modelv2supp1} \\
    \dot{\psiveccomplex}(\xvec,t) &= - \left\{ r-\imag r' -(\diff - \imag\diffp)\nablasquared + \frac{u - \imag u'}{3} \psiveccomplex(\xvec,t)\cdot  
        \psivec(\xvec,t)\right\} \psiveccomplex(\xvec,t) + \sqrt{2\Gamma}\xiveccomplex(\xvec,t) \label{eq:modelv2supp2}
\end{align}
\end{subequations}
To characterize the different phases, it is useful to consider spatially uniform configurations that are the solutions of Eqs.~\eqref{eq:modelv2supp} in the limit of vanishing noise, $\xivec = 0$. Expecting the non-reciprocity to induce global oscillations, we choose the ansatz $\psivec(\xvec,t) = \sqrt{\condensation}e^{-\imag \frequencylc t} \psivec_0$, where $\psivec_0$ is a constant complex vector satisfying $\psiveccomplex_0 \cdot \psivec_0 = 1$. Inserting this into Eqs.~\eqref{eq:modelv2supp} with $\xivec = 0$ and matching the real and imaginary parts, we obtain:
\begin{subequations}
\begin{gather}
    \left( r + \frac{u}{3} \rho_0\right)\sqrt{\rho_0} = 0, \\
    \left(\Omega - r'- \frac{u'}{3} \rho_0\right)\sqrt{\rho_0} = 0.
\end{gather}
\end{subequations}
As usual, existence and stability of the solution depend on the sign of the mass $r$. The stable solutions are given by
\begin{equation} \label{eq:phase}
    (\condensation, \frequencylc) = \begin{cases}
        (0,0), ~\text{for} ~  r >0 \\
        \left( -\frac{3r}{u}, r'- r \frac{u'}{u}\right), ~\text{for} ~ r<0 
    \end{cases}
\end{equation}
where we have used $\rho_0 = -3r/u$ to write $\Omega$ in terms of the parameters of the model. To demonstrate the stability of the oscillatory phase ($r < 0$), we perform a linear stability analysis. Choosing $\psivec(\xvec,t) = \sqrt{\rho_0}\left( \psivec_0 + \wvec(\xvec,t)\right)e^{-\imag \Omega t}$, we derive the linearized equation of motion for $\wvec(\xvec,t)$. Plugging the expansion for $\psivec$ into the Langevin equation in Eq.~\eqref{eq:modelv2supp1}, retaining terms up to the linear order in $\wvec$ and using Eqs.~\eqref{eq:phase} to eliminate terms independent from $\wvec$, we obtain
\begin{align} \label{eq:eomforfluctuations}
    \dot{\wvec}(\xvec,t) &= -\frac{\rho_0}{3}\left( u+\imag u'\right)\psivec_0\left( \psivec_0^{\dagger} \cdot \wvec(\xvec,t) + \wvec^{\dagger}(\xvec,t) \cdot \psivec_0\right) + \left(D+\imag D'\right)\nablasquared\wvec(\xvec,t) \nonumber \\
    &=-|r|\left( 1 +\imag \frac{u'}{u}\right)\psivec_0\left( \psivec_0^{\dagger} \cdot \wvec(\xvec,t) + \wvec^{\dagger}(\xvec,t) \cdot \psivec_0\right) + \left(D+\imag D'\right)\nablasquared\wvec(\xvec,t)
\end{align}
where we have used $\rho_0 = -3r/u$ in going from first to second line. Eq.~\eqref{eq:eomforfluctuations} can be most conveniently analyzed by considering $w(\xvec,t) \equiv \psivec_0^{\dagger} \cdot \wvec(\xvec,t)$ and its complex conjugate $\bar{w}(\xvec,t) \equiv \wvec^{\dagger}(\xvec,t) \cdot \psivec_0 $, the fluctuations of the condensate amplitude and $\wtildevec(\xvec,t) = \left( \boldsymbol{1} - \psivec_0 \psivec_0^{\dagger}\right) \cdot \wvec(\xvec,t)$, the fluctuations in the direction orthogonal to the condensate. Multiplying Eq.~\eqref{eq:eomforfluctuations} by $\boldsymbol{1} - \psivec_0 \psivec_0^{\dagger}$ and using $\psivec_0^{\dagger} \cdot \psivec_0= 1$, we obtain
\begin{equation} \label{eq:eomforgoldstonemode}
    \partial_t\wtildevec(\xvec,t) = \left(D + \imag D'\right)\nablasquared \wtildevec(\xvec,t)
\end{equation}
Due to the diffusive term $D\nablasquared$, all modes decay exponentially in time. Therefore, at long times all fluctuations decay to zero, establishing the linear stability in the transverse direction. To establish stability in the condensate direction, we project Eq.~\eqref{eq:eomforfluctuations} onto $\psivec_0$, giving 
\begin{subequations} \label{eq:eomforfluctuationsparr}
\begin{align}
    \dot{w}(\xvec,t) &= -|r|\left(1 + \imag\frac{u'}{u} \right)\left(w(\xvec,t) + \bar{w}(\xvec,t) \right) + D\left( 1+\imag \frac{D'}{D}\right)\nablasquared w(\xvec,t) \\
    \dot{\bar{w}}(\xvec,t) &= -|r|\left(1 - \imag\frac{u'}{u} \right)\left(w(\xvec,t) + \bar{w}(\xvec,t) \right) + D\left( 1-\imag \frac{D'}{D}\right)\nablasquared \bar{w}(\xvec,t)
\end{align}
\end{subequations}
where we have also included the equation of motion for $\bar{w}(\mathbf{x},t)$, the complex conjugate of $w(\xvec,t)$, for completeness. Analyzing Eq.~\eqref{eq:eomforfluctuationsparr} in Fourier space, one can show that all Fourier modes are linearly stable if and only if $\frac{u'D'}{uD} +  1 >0$ \cite{garcia-morales_complex_2012}, which is satisfied here since $u'D' > 0$ and $u D>0$.

\section{Renormalization Group Procedure} \label{sec:RG}
In this section, we derive the field theory associated with Eqs.~\eqref{eq:modelv2supp}, perform the RG analysis and characterize the dynamics of the model as the system is tuned to criticality. In what follows, we use renormalized field theory with minimal subtraction and dimensional regularization \cite{tauber_critical_2014, zinn-justin_quantum_2002, bellac_quantum_1992, amit_field_2005}. To identify the upper critical dimension, $d_c$, of this model, we perform the usual power counting \cite{tauber_critical_2014}. 
Choosing $\dimx = L$ to be the unit of length, we demand fluctuations and noise remain invariant under spatial rescaling, i.e., $[\Gamma] = B$, $[D] = A$ and $[D'] = A'$, so that the unit of time becomes $[t]= L^2/A = L^2/A'$, which means that $A=A'$. The fields then have dimensions $[\psivec(\xvec,t)] = L^{\frac{2-d}{2}} B^{1/2} A^{-1/2}$ and thus $[u] = [u'] = L^{d-4} A^2 B^{-1}$. For $d > d_c = 4$, these couplings become irrelevant and the critical theory is described by the mean-field exponents. Otherwise, they become relevant. 

\subsection{Field Theory}
In this section, we derive the field theory associated with the Langevin equations \eqref{eq:modelv2supp} and the elements of the perturbative expansion. As a first step, we define the generating functional \cite{janssen_lagrangean_1976, bausch_renormalized_1976, janssen_renormalized_1977, janssen_renormalized_1992,tauber_critical_2014} associated with the Langevin Eqs.~\eqref{eq:modelv2supp}
\begin{equation} \label{eq:Generatingfunc}
    Z[\Jvec, \Jtildevec] = \int \mathcal{D}[\psitildevec, \psivec] \exp{-\action[\psitildevec, \psivec] + \ddint{x} \dint{t} \left( \Jvec^{\dagger} \cdot \psivec + \Jtildevec^{\dagger} \cdot \psitildevec + \rm h.c. \right)}
\end{equation}
where $\rm h.c.$ refers to the Hermitian conjugate of $\Jvec^{\dagger} \cdot \psivec + \Jtildevec^{\dagger} \cdot \psitildevec$. Any correlation function can be calculated by taking the appropriate functional derivatives with respect to the source fields. The action, $\action[\psitildevec, \psivec]$, can be separated into a harmonic part, $\action_0[\psitildevec, \psivec]$
\begin{subequations} \label{eq:action}
\begin{equation} \label{eq:actionharm}
    \action_{0}[\psitildevec, \psivec] = \ddint{x}\dint{t} \left\{\psitildeveccomplex \cdot\left[ (\partial_t + r+ir' - (D+iD')\nablasquared) \psivec - \noise \psitildevec \right] + \psitildevec \cdot\left[ (\partial_t + r-ir' - (D-iD')\nablasquared) \psiveccomplex - \noise \psitildeveccomplex \right] \right\} \,
\end{equation}
and a perturbative part $\action_{I}[\psitildevec, \psivec]$ 
\begin{equation} \label{eq:actionpert}
    \action_{I}[\psitildevec, \psivec] = \ddint{x}\dint{t} \left\{ \frac{u+\imag u'}{3} \left( \psitildeveccomplex \cdot \psivec\right) (\psiveccomplex \cdot \psivec) + \frac{u-\imag u'}{3} \left( \psiveccomplex \cdot \psitildevec\right) (\psiveccomplex \cdot \psivec)\right\}.
\end{equation}
\end{subequations}
Although Eq.~\eqref{eq:modelv2supp2} can be obtained from Eq.~\eqref{eq:modelv2supp1} by complex conjugation, both equations of motion are included in the generating functional. This is because $\psi(\mathbf{x},t)$ and $\psi^\dagger(\mathbf{x},t)$ are treated as independent variables in the field theory \cite{zinn-justin_quantum_2002, tauber_perturbative_2014}. Furthermore, we assign $\tilde{\psivec}^\dagger$ as the response field enforcing Eq.~\eqref{eq:modelv2supp1} and $\tilde{\psivec}$ as the response field enforcing Eq.~\eqref{eq:modelv2supp2}, making the $U(n)$ symmetry manifest: the field theory is invariant under
\begin{subequations}
\begin{align}
    \psivec &\to \unitarymatrix\cdot \psivec &\qquad \psitildevec &\to \unitarymatrix\cdot \psitildevec \\
    \psiveccomplex &\to \psiveccomplex\cdot \unitarymatrixhc &\qquad \psitildeveccomplex &\to \psitildeveccomplex\cdot \unitarymatrixhc
\end{align}
\end{subequations}
for any unitary matrix, $\unitarymatrixhc\cdot \unitarymatrix = \bm{1}$.

To diagonalize the harmonic part of the action Eq.~\eqref{eq:actionharm}, we 
expand the fields in Fourier modes
\begin{subequations} \label{eq:fourierconvention}
    \begin{gather}
        \psivec(\xvec,t) = \ddbarint{k} \dbarint{\omega} e^{\imag\left( \kvec \cdot \xvec - \omega t\right)} \psivec(\kvec,\omega), ~ \psitildevec(\xvec,t) = \ddbarint{k} \dbarint{\omega} e^{\imag\left( \kvec \cdot \xvec - \omega t\right)} \psitildevec(\kvec,\omega) \label{eq:conventionregular}\\
        \psiveccomplex(\xvec,t) = \ddbarint{k} \dbarint{\omega} e^{-\imag\left( \kvec \cdot \xvec - \omega t\right)} \psiveccomplex(\kvec,\omega), ~ \psitildeveccomplex(\xvec,t) = \ddbarint{k} \dbarint{\omega} e^{-\imag\left( \kvec \cdot \xvec - \omega t\right)} \psitildeveccomplex(\kvec,\omega) \label{eq:conventioncc}
    \end{gather}
\end{subequations}
where $\dbar^dk = d^dk/(2\pi)^d$, $\dbar \omega = d\omega/(2\pi)$ and we have chosen to use a different convention to expand Hermitian conjugated fields so that $\psiveccomplex(\kvec, \omega)  = (\psivec(\kvec,\omega))^{\dagger}$ and $\psitildeveccomplex(\kvec, \omega)  = (\psitildevec(\kvec,\omega))^{\dagger}$, consistent with the complex conjugation of the real-space fields. This choice has important consequences for the causality structure of the Feynman diagrams, as we discuss below. With these expansions, the harmonic part of the action in Eq.~\eqref{eq:modelv2supp} diagonalizes as
    \begin{align}
        \action_{0}[\psitildevec, \psivec] &= \ddbarint{k}\dbarint{\omega} \left\{\psitildeveccomplex(\kvec,\omega) \cdot\left[ (-\imag \omega + r+ir' + (D+iD')\kvecsquared) \psivec(\kvec,\omega) - \noise \psitildevec(\kvec,\omega) \right] \right.  \nonumber \\
        & \left.+\psitildevec(\kvec,\omega) \cdot\left[ (\imag \omega + r-ir' +(D-iD')\kvecsquared) \psiveccomplex(\kvec,\omega) - \noise \psitildeveccomplex(\kvec,\omega) \right] \right\} \nonumber \\
        & = \ddbarint{k}\dbarint{\omega} 
        \begin{pmatrix}
            \psitildeveccomplex(\kvec,\omega) & \psiveccomplex(\kvec,\omega)
        \end{pmatrix}
        \cdot
        \begin{pmatrix}
            -2\noise & -\imag \omega + \Delta(\kvec) \\
            \imag\omega + \Deltabar(\kvec) & 0 
        \end{pmatrix}
        \cdot
        \begin{pmatrix}
            \psitildevec(\kvec,\omega) \\ \psivec(\kvec,\omega)
        \end{pmatrix}
    \end{align}
where $\Delta(\kvec) = (D + \imag D')\kvecsquared + r+\imag r'$ and $\Deltabar(\kvec)$ is its complex conjugate. Inverting this matrix yields the bare propagators and the noise vertex
\begin{subequations} \label{eq:diagramsharmonic}
\begin{gather}
  \begin{tikzpicture}[baseline = (o.base)]
  \begin{feynman}
    \vertex (o) at (0,0) {\(a\)};
    \vertex (i) at (2.5,0) {\(b\)};
    \diagram* {
      (o) -- [propagator, edge label={\(\kvec, \omega\)}] (i)
    };
  \end{feynman}
\end{tikzpicture}
 \equiv G_{0}(\kvec, \omega)\kroneckerdelta{} = \frac{\kroneckerdelta{}}{-\imag \omega + \Delta(\kvec)} \\
\begin{tikzpicture}[baseline = (o.base)]
  \begin{feynman}
    \vertex (o) at (0,0) {\(a\)};
    \vertex (i) at (2.5,0) {\(b\)};
    \diagram* {
      (o) -- [anti propagator, edge label={\(\kvec, \omega\)}] (i)
    };
  \end{feynman}
\end{tikzpicture}
\equiv \Gbar_{0}(\kvec, \omega) \kroneckerdelta{} =\frac{\kroneckerdelta{}}{\imag \omega + \Deltabar(\kvec)}, \label{eq:barepropdiagram}\\
\begin{tikzpicture}[baseline = (o.base)]
  \begin{feynman}
    \vertex (o) at (0,0) ;
    \vertex [above left=0.6 of o] (i1) {\(a\)};
    \vertex [below left=0.6 of o] (i2) {\(b\)};
    \diagram* {
      (o) -- [scalar] (i1),
      (o) -- [plain] (i2)
    };
  \end{feynman}
\end{tikzpicture}
=2\noise\kroneckerdelta{}. \label{eq:noisevertexdiagram}
\end{gather}
\end{subequations}
where we have defined the bare propagator $G_0(\kvec, \omega) = 1/(-\imag \omega + \Delta(\kvec))$, removed the arrows from the external legs of the amputated vertices and used solid lines to denote $\psivec$ and $\tilde{\psivec}$, and dashed lines to denote the complex conjugate fields $\psivec^\dagger$ and $\tilde{\psivec}^\dagger$. Generally, in dynamical field theories, all arrows flow in a fixed certain direction due to causality, typically chosen from left to right \cite{tauber_critical_2014}. This is a direct consequence of the propagator having a simple pole at lower half of the complex $\omega$-plane. However, in the present work, we have a propagator, $\Gbar_0(\kvec, \omega)$, whose pole lies in the upper half of the complex plane. This reverses its causal structure; consequently, $\overline{G}_0$ can be interpreted as propagating `backward' in time. This is analogous to antiparticles in quantum field theory, which are interpreted as propagating backwards in time \cite{zinn-justin_quantum_2002}. In the present case, we adopt the convention that arrows always point toward solid lines. For example, at the tree-level, the correlation function is given by:
\begin{equation} \label{eq:correlationfunctiontreelevel}
\begin{tikzpicture}[baseline = (o.base)]
  \begin{feynman}
    \vertex (o) at (0,0) ;
    \vertex [above left=of o] (i1) {\(a\)};
    \vertex [below left=of o] (i2) {\(b\)};
    \diagram* {
      (o) -- [anti propagator, edge label'={\(\kvec, \omega\)}] (i1),
      (o) -- [propagator, edge label={\(\kvec, \omega\)}] (i2)
    };
  \end{feynman}
\end{tikzpicture}
\equiv C_0(\kvec,\omega) \kroneckerdelta{} = \frac{2\noise \kroneckerdelta{}}{\left( -\imag \omega + \Delta(\kvec)\right)\left( \imag \omega + \Deltabar(\kvec)\right)}. 
\end{equation}
To find the interaction vertices, we rewrite Eq.~\eqref{eq:actionpert} in terms of the Fourier transformed fields.  This gives us the following expression:
\begin{subequations} \label{eq:actionpertfourier}
\begin{align}
    \action_{I}[\psitildevec, \psivec] = \int \prod_{i = 1}^4 \dbar^d k_i \dbar\omega_i \deltabar\left( \sum_{i = 1}^4 \kvec_i\right) \deltabar\left( \sum_{i = 1}^4 \omega_i\right) &\left\{ \frac{u+\imag u'}{6}\Qtensor{} \psitildeveccomplexcomp[a](-\kvec_1, -\omega_1) \psiveccomplexcomp[b](-\kvec_2,-\omega_2) \psiveccomp[c](\kvec_3, \omega_3) \psiveccomp[d](\kvec_4,\omega_4) \right. \nonumber \\
    &\left.+ \frac{u-\imag u'}{6} \Qtensor{}\psitildeveccomp[a](\kvec_1, \omega_1) \psiveccomp[b](\kvec_2,\omega_2) \psiveccomplexcomp[c](-\kvec_3, -\omega_3) \psiveccomplexcomp[d](-\kvec_4,-\omega_4)\right\}.
\end{align}
\end{subequations}
where $\Qtensor{} = \kroneckerdelta{ac}\kroneckerdelta{bd} + \kroneckerdelta{ad} \kroneckerdelta{bc}$ and the bar denotes complex conjugation. In Eq.~\eqref{eq:actionpertfourier}, we employ the Einstein summation convention, summing over repeated indices, and have symmetrized the action over field components. From Eq.~\eqref{eq:actionpertfourier}, we can read off the perturbative vertices:
\begin{equation}
    \begin{tikzpicture}[baseline = (o.base), scale = 1.2]
  \begin{feynman}
    \vertex (origin) at (0,0) ;
    \vertex (o) at (-0.6,0) {\(a\)};
    \vertex (i2) at (0.6,0) {\(c\)};
    \vertex (i1) at (0.6,0.6) {\(b\)};
    \vertex (i3) at (0.6,-0.6) {\(d\)};
    \diagram* {
      (origin) -- [scalar] (o),
      (origin) -- [scalar] (i1),
      (origin) -- [plain] (i2),
      (origin) -- [plain] (i3),
    };
  \end{feynman}
\end{tikzpicture}
=-\frac{u+\imag u'}{3} \Qtensor{},  ~~
\begin{tikzpicture}[baseline = (o.base), scale = 1.2]
  \begin{feynman}
    \vertex (origin) at (0,0) ;
    \vertex (o) at (-0.6,0) {\(a\)};
    \vertex (i2) at (0.6,0) {\(c\)};
    \vertex (i1) at (0.6,0.6) {\(b\)};
    \vertex (i3) at (0.6,-0.6) {\(d\)};
    \diagram* {
      (origin) -- [plain] (o),
      (origin) -- [plain] (i1),
      (origin) -- [scalar] (i2),
      (origin) -- [scalar] (i3),
    };
  \end{feynman}
\end{tikzpicture}
=-\frac{u-\imag u'}{3}\Qtensor{},\label{eq:speciescouplingvertices}
\end{equation}
where the amputated vertices are symmetric under the individual exchange of incoming solid and dashed lines.

\subsection{Renormalization Scheme}
In this section, we outline our renormalization scheme. Perturbative RG is carried out most elegantly by considering \emph{vertex functions} associated with the correlation functions \cite{tauber_critical_2014, amit_field_2005, bellac_quantum_1992}. They are defined to be
\begin{subequations} \label{eq:vertexfuncdefn}
\begin{gather}
    \invprop{}{} = \frac{\kroneckerdelta{}}{G(\kvec,\omega)}, ~~\invpropcomplex{}{} = \frac{\kroneckerdelta{}}{\Gbar(\kvec,\omega)}, \\
        \vertexnoise{}{\kvec,\omega} = \kroneckerdelta{} \frac{C(\kvec,\omega)}{G(\kvec,\omega) \Gbar(\kvec,\omega)}, \\
        \vertexnonlin{}{\kvec_2, \omega_2, \kvec_3, \omega_3, \kvec_4, \omega_4} \deltabar(-\kvec_1-\kvec_2+\kvec_3 + \kvec_4)\deltabar(-\omega_1-\omega_2 + \omega_3 + \omega_4) = \nonumber \\
        \frac{\av{\psiveccomp[a](\kvec_1, \omega_1) \psitildeveccomp[b](\kvec_2,\omega_2) \psitildeveccomplexcomp[c](\kvec_3,\omega_3) \psitildeveccomplexcomp[d](\kvec_4,\omega_4)}_c}{G(\kvec_1, \omega_1) \Gbar(\kvec_2, \omega_2) G(\kvec_3, \omega_3) G(\kvec_4, \omega_4)}  \\
        \vertexnonlincomplex{}{\kvec_2, \omega_2, \kvec_3, \omega_3, \kvec_4, \omega_4} \deltabar(\kvec_1+\kvec_2-\kvec_3 - \kvec_4)\deltabar(\omega_1+\omega_2 - \omega_3 - \omega_4) = \nonumber \\
        \frac{\av{\psiveccomplexcomp[a](\kvec_1, \omega_1) \psitildeveccomplexcomp[b](\kvec_2,\omega_2) \psitildeveccomp[c](\kvec_3,\omega_3) \psitildeveccomp[d](\kvec_4,\omega_4)}_c}{\Gbar(\kvec_1, \omega_1) G(\kvec_2, \omega_2) \Gbar(\kvec_3, \omega_3) \Gbar(\kvec_4, \omega_4)}
\end{gather}
\end{subequations}
where we have defined the full propagators $\langle \psiveccomp[a](\kvec,\omega) \psitildeveccomplexcomp[b](\kvec',\omega')\rangle = \kroneckerdelta{}\deltabar(\kvec-\kvec') \deltabar(\omega - \omega')G(\kvec,\omega)$ and $\langle \psiveccomplexcomp[a](\kvec,\omega) \psitildeveccomp[b](\kvec',\omega')\rangle = \kroneckerdelta{}\deltabar(\kvec-\kvec') \deltabar(\omega - \omega')\Gbar(\kvec,\omega)$ and $c$ denotes the connected part of the correlation function. At tree-level, the vertex functions can be read-off from the action Eq.~\eqref{eq:action}
\begin{subequations} \label{eq:treelevelvertex}
    \begin{gather}
        \invprop{}{} = \kroneckerdelta{}\left( -\imag \omega +r+ir' + (D+iD')\kvecsquared \right) + \mathcal{O}(\lambda), ~~\invpropcomplex{}{} = \kroneckerdelta{}\left( \imag \omega +r-ir' + (D-iD')\kvecsquared \right) + \mathcal{O}(\lambda), \\
        \vertexnoise{}{\kvec,\omega} = 2\noise \kroneckerdelta{} + \mathcal{O}(\lambda), \\
        \vertexnonlin{}{\kvec_2, \omega_2, \kvec_3, \omega_3, \kvec_4, \omega_4} = -\frac{u + \imag u'}{3}\Qtensor{} + \mathcal{O}(\lambda^2), ~ \vertexnonlincomplex{}{\kvec_2, \omega_2, \kvec_3, \omega_3, \kvec_4, \omega_4} = -\frac{u - \imag u'}{3}\Qtensor{} + \mathcal{O}(\lambda^2)
    \end{gather}
\end{subequations}
where $\mathcal{O}(\lambda)$ denotes corrections of order $u$ or $u'$. At tree-level, some of the vertices are related to each other through complex conjugation, $\invprop{}{} = \left( \invpropcomplex{}{}\right)^*$ and $\vertexnonlin{}{\kvec_2, \omega_2, \kvec_3, \omega_3, \kvec_4, \omega_4} = \left( \vertexnonlincomplex{}{\kvec_2, \omega_2, \kvec_3, \omega_3, \kvec_4, \omega_4}\right)^{*}$. These relations will still hold once we include diagrammatic corrections to these vertices as follows directly by using the definition of the vertices in Eq.\eqref{eq:vertexfuncdefn}. Therefore, to carry out the RG procedure, we only have to consider the corrections to $\invprop{}{}, \vertexnoise{}{\kvec,\omega}$ and $\vertexnonlin{}{\kvec_2, \omega_2, \kvec_3, \omega_3, \kvec_4, \omega_4}$. Furthermore, because we are only interested in the RG \emph{relevant} corrections to the vertices, we can safely set the external wave-vector and frequency to be zero in diagrams renormalizing $\vertexnoise{}{\kvec,\omega}$ and $\vertexnonlin{}{\kvec_1, \omega_1; \kvec_2, \omega_2, \kvec_3, \omega_3, \kvec_4, \omega_4}$ as the wave-vector or frequency dependent parts of these vertices have no bearing on the critical behavior of the theory. Finally, for notational ease, we use the convention
\begin{subequations}
    \begin{align}
        \vertexnoise{}{} &\defequal \vertexnoise{}{\zerovec,0} \\
        \vertexnonlin{}{} &\defequal \vertexnonlin{}{\zerovec, 0, \zerovec, 0, \zerovec, 0}
    \end{align}
\end{subequations}

To perform the RG procedure, we first identify the critical point of the theory where the correlation lengths diverge. At mean-field level, this is given by $r_{c}=r'_{c}=0$. However, as is well understood \cite{tauber_critical_2014}, this value will be shifted by the fluctuations of the fields. The critical point $(r_{c}, r'_{c})$ is then determined by the simultaneous roots of:
\begin{equation}\label{eq:criticalpoint}
    \left. \invprop{}{\zerovec,0} \right|_{r = r_c, r'=r'_c} = 0,
\end{equation}
ensuring the field to have infinite correlation length. Through a suitable field transformation, $r'$ can always be tuned to its critical value $r_c'$, or simply gauged away, meaning the critical behavior is then strictly controlled by $r$ alone. Although removing $r'$ by a field redefinition might seem to simplify the RG analysis, it merely transfers its renormalization to the fields \cite{risler_universal_2004} through time-dependent renormalization constants, unnecessarily complicating the procedure. We therefore find it more convenient to retain $r'$. Upon solving Eq.~\eqref{eq:criticalpoint} for $r_c$ and $r'_c$, we can start parameterizing the system by $\tau = r-r_c$ and $\tau'= r'-r'_c$, the distances away from the critical point. We can then proceed with eliminating the divergences that arise in the perturbative expansion. This elimination is done at an arbitrary inverse-length scale $\mu$. To absorb the UV-divergences that arise in the perturbative expansion, we introduce the following $Z$-factors and renormalized quantities, indicated by the subscript $R$: First, the field renormalizations
\begin{subequations} \label{eq:Zfactorsdefn}
    \begin{equation} \label{eq:Zfactorsfield}
        \psivec_R = \Zpsi \psivec, \qquad\psitildevec_R = \Zpsitilde\psitildevec,
    \end{equation}
and second, the renormalization of the couplings
\begin{gather}  \label{eq:Zfactorscoupling}
     \massR = \Zmass \tau \mu^{-2}, ~~\masspR = \Zmassp \taup \mu^{-2}, \\
    \diffR = \ZDiff D, ~~\diffpR = \ZDiffp \diffp,~ ~\nonlincouplingR = \Znonlincoupling\nonlincoupling\mueps{-}, ~~\nonlincouplingpR = \Znonlincouplingp \nonlincouplingp \mueps{-}.
\end{gather}
\end{subequations}
Any renormalized coupling has been made spatially dimensionless using $\epsilon = 4- d$ for the nonlinearities. With this adjustment, renormalized couplings remain finite and any physical observable will be given in terms of the renormalized quantities which are the well-defined objects in the field theory. Since the fields $\psivec$ and $\psitildevec$ are complex variables, we expect their respective $Z-$factors to be complex as well. 

To carry out the renormalization procedure, we find it convenient to define the dimensionless expansion parameters 
\begin{equation} \label{eq:defexpansionparameters}
    \utilde \defequal \frac{u \noise}{(4\pi)^{d/2} D^2}, \quad \rationonlin \defequal \frac{u'}{u}, \quad \ratiodiff \defequal \frac{D'}{D}
\end{equation}
and their renormalized counterparts $\utildeR = Z_{\utilde} \utilde  \mu^{-\epsilon}$,  $\rationonlinR \equiv Z_{\rationonlin}\rationonlin$ and $\ratiodiffR \equiv Z_{\ratiodiff}\ratiodiff$. The $Z-$factors for these parameters can be expressed in terms of $Z-$factors in Eq.~\eqref{eq:Zfactorsdefn}
\begin{equation}
    Z_{\utilde} = \frac{Z_{u} Z_{\noise}}{Z^2_{D}}, \quad Z_{\rationonlin} = \frac{\Znonlincouplingp}{\Znonlincoupling}, \quad \ZwD = \frac{\ZDiffp}{\ZDiff}
\end{equation}

The precise choice of renormalization scheme is arbitrary. We choose the normalization point (NP) at finite $\tau_R$ and $\tau'_R$, with the critical point approached as $\mu \to 0$. Furthermore, adhering to the minimal subtraction scheme \cite{bellac_quantum_1992}, we impose the following normalization conditions on the renormalized vertices, to leading order in $1/\epsilon$ divergences
\begin{subequations} \label{eq:renormconds}
    \begin{align}
        \left. \invpropren{}{} \right|_{\rm NP} &= \kroneckerdelta{}\left( -\imag \omega + \mu^2(\tau_R+\imag\tau_R') + (D_R+\imag D'_R) \kvecsquared\right) \\
        \left. \vertexnoiseren{}{} \right|_{\rm NP} &= 2\noiseR \kroneckerdelta{} \\
        \left. \vertexnonlinren{}{} \right|_{\rm NP} &= -\frac{u_R + \imag u'_R}{3}\mueps{}\Qtensor{}
    \end{align}
\end{subequations}
where the renormalized vertices are defined to be 
\begin{subequations} \label{eq:renvertexfuncdefn}
\begin{gather}
    \invpropren{}{} = \frac{\kroneckerdelta{}}{G_R(\kvec,\omega)}, \\
        \vertexnoiseren{}{\kvec,\omega} = \kroneckerdelta{} \frac{C_R(\kvec,\omega)}{G_R(\kvec,\omega) \Gbar_R(\kvec,\omega)}, \\
        \vertexnonlinren{}{\kvec_2, \omega_2, \kvec_3, \omega_3, \kvec_4, \omega_4} \deltabar(-\kvec_1-\kvec_2+\kvec_3 + \kvec_4)\deltabar(-\omega_1-\omega_2 + \omega_3 + \omega_4) = \\
        \frac{\av{\psiveccomp[a]_R(\kvec_1, \omega_1) \psitildeveccomp[b]_R(\kvec_2,\omega_2) \psitildeveccomplexcomp[c]_R(\kvec_3,\omega_3) \psitildeveccomplexcomp[d]_R(\kvec_4,\omega_4)}_c}{G_R(\kvec_1, \omega_1) \Gbar_R(\kvec_2, \omega_2) G_R(\kvec_3, \omega_3) G_R(\kvec_4, \omega_4)}
\end{gather}
\end{subequations}
with $\langle \psiveccomp[a]_R(\kvec, \omega) \psitildeveccomplexcomp[b]_R(\kvec', \omega')\rangle = \kroneckerdelta{} \deltabar(\kvec-\kvec') \deltabar(\omega -\omega') G_{R}(\kvec, \omega)$ and $\langle \psiveccomplexcomp[a]_R(\kvec, \omega) \psitildeveccomp[b]_R(\kvec', \omega')\rangle = \kroneckerdelta{} \deltabar(\kvec-\kvec') \deltabar(\omega -\omega') \Gbar_{R}(\kvec, \omega)$. Using the definition of the renormalized fields in Eq.~\eqref{eq:Zfactorsdefn}, we have
\begin{subequations} \label{eq:renvertextonormalvertex}
\begin{gather}
       \invpropren{}{} = \Zpsitildecomplex^{-1} \Zpsi^{-1}\invprop{}{}, \\
        \vertexnoiseren{}{\kvec,\omega} = |\Zpsitilde|^{-2}\vertexnoise{}{\kvec,\omega}, \\
        \vertexnonlinren{}{\kvec_2, \omega_2, \kvec_3, \omega_3, \kvec_4, \omega_4}  = \Zpsitildecomplex^{-1} \Zpsicomplex^{-1} \Zpsi^{-2}\vertexnonlin{}{\kvec_2, \omega_2, \kvec_3, \omega_3, \kvec_4, \omega_4}
\end{gather}
\end{subequations}
The conditions in Eqs.~\eqref{eq:renormconds} are sufficient to fix the values of the $Z-$factors. In the next section, we use Eq.~\eqref{eq:renormconds} along with \eqref{eq:renvertextonormalvertex} to find these $Z-$ factors.

\subsection{Loop Calculation}
In this section, we outline the details of the loop calculation. We begin by identifying the one-loop critical point $r_c,r'_c$. To this end, we consider all one-loop diagrams contributing to the inverse propagator $\invprop{}{}$:
\begin{align} \label{eq:invproponeloop}
    \invprop{}{} &= \left(-\imag \omega + r + \imag r'+ \left( D + \imag D' \right)\kvecsquared \right) \kroneckerdelta{} - 
    \begin{tikzpicture}[baseline = (o.base), scale = 1.2]
   \begin{feynman}
    \vertex (o) at (0,0) {\(a\)};
    \vertex (a) at (0.5,0);
    \vertex (b) at (1.5,1);;
    \vertex (i) at (1,0) {\(b\)};
    \diagram* {
      (a) -- [anti propagator, bend left=20] (b),
      (a) -- [propagator, bend right=20] (b),
      (o) -- [scalar ] (a),
      (a) -- [plain] (i)
    };
  \end{feynman}
\end{tikzpicture} \nonumber
\\
&= \left(-\imag \omega + r + \imag r'+ \left( D + \imag D' \right)\kvecsquared \right) \kroneckerdelta{} +  \frac{\numcomponent + 1}{3} u(1+\imag w_u) \ddbarint{q_1} \dbarint{\omega_1}\frac{2 \noise}{(-\imag \omega_1 + \Delta(\qvec_1)) (\imag \omega_1 + \Deltabar(\qvec_1)) }
\end{align}
where the extra minus sign in the first line arises from the geometric summation of the amputated diagram. Rather than performing an additive renormalization \cite{tauber_critical_2014}, we replace $r$ by $\tau= \mu^2\tau_{R}+\mathcal{O}(\lambda)$ and $r'= \tau'_R\mu^2 + \mathcal{O}(\lambda^2)$, thereby regularizing the infrared (IR). Where we encounter IR divergences of the form $\tau^{-\epsilon/2}$ or $\tau'^{-\epsilon/2}$, we may replace them by $\mu^{-\epsilon/2}$ to leading order in small $\epsilon$ as $\tau_{R}$ and $\tau'_R$ are kept finite. As for the ultraviolet (UV), we use the analytic continuation of the UV-divergence captured by $\Gamma(1-d/2)=-\Gamma(\epsilon/2)+\mathcal{O}(\epsilon^0)$ to write \Eref{invproponeloop} as
\begin{align} \label{eq:invprop_oneloop_algebraic}
    \invprop{}{} = \left[-\imag \omega + (D+ \imag D')\kvecsquared\right] \kroneckerdelta{}& +\left[\tau\left( 1 - \frac{\numcomponent +1}{3} \utilde \mueps{-} \frac{\Gamma(\epsilon/2)}{(4 \pi)^{d/2}} \right)  + \imag \tau'\left( 1 - \frac{\numcomponent +1}{3} \utilde \mueps{-}\rationonlin \frac{\tau}{\taup} \frac{\Gamma(\epsilon/2)}{(4 \pi)^{d/2}} \right) \right] \kroneckerdelta{}.
\end{align}
where $\Gamma(z)$ is the Euler Gamma function and 
the wave-vector integrals are evaluated using the identity \cite{tauber_critical_2014}:
\begin{equation} \label{eq:integral_identity}
 \ddbarint{k} \frac{|\kvec|^{2\sigma}}{(\kvecsquared + \tau)^s} = \frac{\Gamma(d/2 + \sigma) \Gamma(s-\sigma-d/2)}{\forpidtwo \Gamma(d/2)\Gamma(s)} \tau^{\sigma - s+d/2}.    
\end{equation} 
By applying the renormalization conditions, Eqs.~\eqref{eq:renormconds} on Eq.~\eqref{eq:invprop_oneloop_algebraic} and using Eqs.~\eqref{eq:renvertextonormalvertex}, we find 
\begin{equation} \label{eq:zfactormass}
    \Zpsi \Zpsitildecomplex = \ZDiff=\ZDiffp = 1,~\Zmass = 1-\frac{2(\numcomponent + 1)}{3}  \frac{ \utilde \mueps{-}}{\epsilon} , ~\Zmassp = 1-\frac{2(\numcomponent + 1)}{3}  \rationonlin\frac{\tau}{\taup} \frac{\utilde \mueps{-}}{\epsilon}
\end{equation}
where we have used the identity $\Gamma(\epsilon) \sim 1/\epsilon + \mathcal{O}(\epsilon^0)$. The $Z$-factor for $\tau'$ depends on the ratio $\tau'/\tau$, which might flow under RG. However, as we explain below, this flow need not be captured. As is common in field theories with a quartic nonlinearity, $Z_{D}$ and field renormalizations do not receive corrections at one-loop level. 

We proceed with the renormalization of the noise vertex. To one-loop, it does not receive any renormalization. Therefore, the $Z$-factors can be found to satisfy:
\begin{equation}
    \Znoise |\Zpsitilde|^2 = 1 + \mathcal{O}(u^2)
\end{equation}

Finally, we consider the renormalization of the nonlinear vertex $\vertexnonlin{}{}$ to one-loop
\begin{equation}
    \vertexnonlin{}{} = 
    \begin{tikzpicture}[baseline = (o.base), scale = 1.2]
  \begin{feynman}
    \vertex (origin) at (0,0) ;
    \vertex (o) at (-0.6,0) {\(a\)};
    \vertex (i2) at (0.6,0) {\(c\)};
    \vertex (i1) at (0.6,0.6) {\(b\)};
    \vertex (i3) at (0.6,-0.6) {\(d\)};
    \diagram* {
      (origin) -- [scalar] (o),
      (origin) -- [scalar] (i1),
      (origin) -- [plain] (i2),
      (origin) -- [plain] (i3),
    };
  \end{feynman}
\end{tikzpicture}
+ 
\begin{tikzpicture}[baseline = (o.base), scale = 1.2]
\begin{feynman}
\vertex (o) at (0,0) {\(a\)};
\vertex (a) at (0.6,0);
\vertex (i3) at ($(a) + (0.6,-0.5)$) {\(b\)};
\vertex (b) at ($(a) + (0.33,1)$);     
\vertex (c) at ($(a) + (1,0.4)$);    
\vertex (i1) at ($(b) + (0.5,0.6)$) {\(c\)};  
\vertex (i2) at ($(b) + (0.5,0.3)$) {\(d\)};  
\diagram* {
(o) -- [scalar] (a),
(a) -- [scalar] (i3),
(a) -- [propagator, edge label = {\(-\qvec_1, -\omega_1\)}] (b),
(b) -- [anti propagator, edge label = {\(\qvec_1, \omega_1\)}] (c),
(a) -- [propagator, edge label' = {\(\qvec_1, \omega_1\)}] (c),
(b) -- [plain] (i1),
(b) -- [plain] (i2),
};
\end{feynman}
\end{tikzpicture}
~~~ + \text{sym.} 
~+~
\begin{tikzpicture}[baseline = (o.base), scale = 1.2]
\begin{feynman}
\vertex (o) at (0,0) {\(a\)};
\vertex (a) at (0.6,0);
\vertex (i3) at ($(a) + (0.6,-0.5)$) {\(d\)};
\vertex (b) at ($(a) + (0.33,1)$);     
\vertex (c) at ($(a) + (1,0.4)$);    
\vertex (i1) at ($(b) + (0.5,0.6)$) {\(b\)};  
\vertex (i2) at ($(b) + (0.5,0.3)$) {\(c\)};  
\diagram* {
(o) -- [scalar] (a),
(a) -- [plain] (i3),
(a) -- [anti propagator, edge label = {\(\qvec_1, \omega_1\)}] (b),
(b) -- [anti propagator, edge label = {\(\qvec_1, \omega_1\)}] (c),
(a) -- [propagator, edge label' = {\(\qvec_1, \omega_1\)}] (c),
(b) -- [scalar] (i1),
(b) -- [plain] (i2),
};
\end{feynman}
\end{tikzpicture}
~~~ + \text{sym.} +
\begin{tikzpicture}[baseline = (o.base),scale = 1.2]
\begin{feynman}
\vertex (o) at (0,0) {\(a\)};
\vertex (a) at (0.6,0);
\vertex (i3) at ($(a) + (0.6,-0.5)$) {\(d\)};
\vertex (b) at ($(a) + (0.33,1)$);     
\vertex (c) at ($(a) + (1,0.4)$);    
\vertex (i1) at ($(b) + (0.5,0.6)$) {\(b\)};  
\vertex (i2) at ($(b) + (0.5,0.3)$) {\(c\)};  
\diagram* {
(o) -- [scalar] (a),
(a) -- [plain] (i3),
(a) -- [propagator, edge label = {\(\qvec_1, \omega_1\)}] (b),
(b) -- [propagator, edge label = {\(\qvec_1, \omega_1\)}] (c),
(a) -- [anti propagator, edge label' = {\(\qvec_1, \omega_1\)}] (c),
(b) -- [scalar] (i1),
(b) -- [plain] (i2),
};
\end{feynman}
\end{tikzpicture} 
+ \text{sym.}
\end{equation}
where sym. refers to the symmetrization of the diagrams over the incoming legs, such as 
\[
\begin{tikzpicture}[baseline = (o.base), scale = 1.2]
\begin{feynman}
\vertex (o) at (0,0) {\(a\)};
\vertex (a) at (0.6,0);
\vertex (i3) at ($(a) + (0.6,-0.5)$) {\(d\)};
\vertex (b) at ($(a) + (0.33,1)$);     
\vertex (c) at ($(a) + (1,0.4)$);    
\vertex (i1) at ($(b) + (0.5,0.6)$) {\(b\)};  
\vertex (i2) at ($(b) + (0.5,0.3)$) {\(c\)};  
\diagram* {
(o) -- [scalar] (a),
(a) -- [plain] (i3),
(a) -- [anti propagator, edge label = {\(\qvec_1, \omega_1\)}] (b),
(b) -- [anti propagator, edge label = {\(\qvec_1, \omega_1\)}] (c),
(a) -- [propagator, edge label' = {\(\qvec_1, \omega_1\)}] (c),
(b) -- [scalar] (i1),
(b) -- [plain] (i2),
};
\end{feynman}
\end{tikzpicture}
~+~
\begin{tikzpicture}[baseline = (o.base), scale = 1.2]
\begin{feynman}
\vertex (o) at (0,0) {\(a\)};
\vertex (a) at (0.6,0);
\vertex (i3) at ($(a) + (0.6,-0.5)$) {\(c\)};
\vertex (b) at ($(a) + (0.33,1)$);     
\vertex (c) at ($(a) + (1,0.4)$);    
\vertex (i1) at ($(b) + (0.5,0.6)$) {\(b\)};  
\vertex (i2) at ($(b) + (0.5,0.3)$) {\(d\)};  
\diagram* {
(o) -- [scalar] (a),
(a) -- [plain] (i3),
(a) -- [anti propagator, edge label = {\(\qvec_1, \omega_1\)}] (b),
(b) -- [anti propagator, edge label = {\(\qvec_1, \omega_1\)}] (c),
(a) -- [propagator, edge label' = {\(\qvec_1, \omega_1\)}] (c),
(b) -- [scalar] (i1),
(b) -- [plain] (i2),
};
\end{feynman}
\end{tikzpicture}
\]
We evaluate the diagrams one-by-one, taking care of subtleties regarding the causality
\begin{subequations}
    \begin{align}
        \begin{tikzpicture}[baseline = (o.base), scale = 1.2]
\begin{feynman}
\vertex (o) at (0,0) {\(a\)};
\vertex (a) at (0.6,0);
\vertex (i3) at ($(a) + (0.6,-0.5)$) {\(b\)};
\vertex (b) at ($(a) + (0.33,1)$);     
\vertex (c) at ($(a) + (1,0.4)$);    
\vertex (i1) at ($(b) + (0.5,0.6)$) {\(c\)};  
\vertex (i2) at ($(b) + (0.5,0.3)$) {\(d\)};  
\diagram* {
(o) -- [scalar] (a),
(a) -- [scalar] (i3),
(a) -- [propagator, edge label = {\(-\qvec_1, -\omega_1\)}] (b),
(b) -- [anti propagator, edge label = {\(\qvec_1, \omega_1\)}] (c),
(a) -- [propagator, edge label' = {\(\qvec_1, \omega_1\)}] (c),
(b) -- [plain] (i1),
(b) -- [plain] (i2),
};
\end{feynman}
\end{tikzpicture} + \text{sym.}
& = \Qtensor{}\frac{2}{9} u^2(1+\imag \rationonlin)^2 \ddbarint{q_1}\dbarint{\omega_1} \frac{2\noise}{(\imag \omega_1 + \Delta(-\qvec_1))(\imag \omega_1 + \Deltabar(\qvec_1))(-\imag \omega_1 + \Delta(\qvec_1))} \nonumber \\
& = \Qtensor{}\frac{2}{9}u^2(1+\imag \rationonlin)^2 \Gamma \ddbarint{q_1} \frac{1}{\Delta(\qvec_1)(\Delta(\qvec_1) + \Deltabar(\qvec_1))} \nonumber \\
&= \Qtensor{}\frac{1}{9}\frac{u^2 \noise}{D^2}\frac{(1+\imag \rationonlin)^2}{1+\imag \ratiodiff} \ddbarint{q_1} \frac{1}{(\qvec_1^2 +\tau/D)(\qvec_1^2 + (\tau + \taup)/(D + \imag D'))} \\
\begin{tikzpicture}[baseline = (o.base), scale = 1.2]
\begin{feynman}
\vertex (o) at (0,0) {\(a\)};
\vertex (a) at (0.6,0);
\vertex (i3) at ($(a) + (0.6,-0.5)$) {\(d\)};
\vertex (b) at ($(a) + (0.33,1)$);     
\vertex (c) at ($(a) + (1,0.4)$);    
\vertex (i1) at ($(b) + (0.5,0.6)$) {\(b\)};  
\vertex (i2) at ($(b) + (0.5,0.3)$) {\(c\)};  
\diagram* {
(o) -- [scalar] (a),
(a) -- [plain] (i3),
(a) -- [anti propagator, edge label = {\(\qvec_1, \omega_1\)}] (b),
(b) -- [anti propagator, edge label = {\(\qvec_1, \omega_1\)}] (c),
(a) -- [propagator, edge label' = {\(\qvec_1, \omega_1\)}] (c),
(b) -- [scalar] (i1),
(b) -- [plain] (i2),
};
\end{feynman}
\end{tikzpicture} 
+ \text{sym .}& = \Qtensor{} \frac{\numcomponent + 3}{9} u^2(1 + \rationonlin^2) \ddbarint{q_1} \dbarint{\omega_1} \frac{2\noise}{(-\imag \omega_1 + \Delta(\qvec_1))(\imag \omega_1 + \Deltabar(\qvec_1))^2} \nonumber \\
& = \Qtensor{}\frac{2(\numcomponent+3)}{9}u^2(1+\rationonlin^2)\noise \ddbarint{q_1} \frac{1}{(\Delta(\qvec_1) + \Deltabar(\qvec_1))^2} \nonumber \\
&= \Qtensor{}\frac{\numcomponent+3}{18}\frac{u^2 \noise}{D^2}(1+\rationonlin^2) \ddbarint{q_1} \frac{1}{(\qvecsquared_1 + \tau/D)^2} \\
\begin{tikzpicture}[baseline = (o.base),scale = 1.2]
\begin{feynman}
\vertex (o) at (0,0) {\(a\)};
\vertex (a) at (0.6,0);
\vertex (i3) at ($(a) + (0.6,-0.5)$) {\(d\)};
\vertex (b) at ($(a) + (0.33,1)$);     
\vertex (c) at ($(a) + (1,0.4)$);    
\vertex (i1) at ($(b) + (0.5,0.6)$) {\(b\)};  
\vertex (i2) at ($(b) + (0.5,0.3)$) {\(c\)};  
\diagram* {
(o) -- [scalar] (a),
(a) -- [plain] (i3),
(a) -- [propagator, edge label = {\(\qvec_1, \omega_1\)}] (b),
(b) -- [propagator, edge label = {\(\qvec_1, \omega_1\)}] (c),
(a) -- [anti propagator, edge label' = {\(\qvec_1, \omega_1\)}] (c),
(b) -- [scalar] (i1),
(b) -- [plain] (i2),
};
\end{feynman}
\end{tikzpicture}
+ \text{sym.} &= \Qtensor{} \frac{\numcomponent+3}{9}u^2(1+\imag \rationonlin)^2 \ddbarint{q_1} \dbarint{\omega_1} \frac{2 \noise}{(-\imag \omega_1 + \Delta(\qvec_1))^2(\imag \omega_1 +\Deltabar(\qvec_1))} \nonumber \\
&= \Qtensor{} \frac{\numcomponent+3}{18} \frac{u^2 \noise}{D^2}(1+\imag \rationonlin)^2 \ddbarint{q_1} \frac{1}{(\qvec_1^2 + \tau/D)^2}
    \end{align}
\end{subequations}
Evaluating these integrals at the normalization point, we obtain the following vertex at the NP:
\begin{equation} \label{eq:quartic_vertex_NP}
    \left. \vertexnonlin{}{} \right|_{\rm NP} = -\frac{\Qtensor{}}{3}u(1+\imag \rationonlin)\left( 1 - \frac{2}{3}\frac{1+\imag \rationonlin}{1+\imag \ratiodiff} \frac{\utilde\mueps{-}}{\epsilon}  - \frac{2(\numcomponent+3)}{3}\frac{\utilde \mueps{-}}{\epsilon} \right)
\end{equation}
Applying the renormalization condition in Eq.~\eqref{eq:renormconds} and using Eq.~\eqref{eq:renvertextonormalvertex}, we can find the $Z$-factors associated with $u$ and $u'$ by separating Eq.~\eqref{eq:quartic_vertex_NP} into real and imaginary parts:
\begin{equation} \label{eq:zfactorsnonlincoupling}
        \Znonlincoupling = 1 + \frac{2}{3}\frac{(\rationonlin - \ratiodiff)^2}{1+w^2_D} \frac{\utilde \mueps{-}}{\epsilon} - \frac{2(\numcomponent+4)}{3} \frac{\utilde \mueps{-}}{\epsilon}, ~\Zwu= 1 - \frac{2}{3}\frac{(\rationonlin-\ratiodiff)(1+\rationonlin^2)}{(1+\ratiodiff^2)\rationonlin}\frac{\utilde \mueps{-}}{\epsilon}.
\end{equation}
The $Z$-factors obtained thus far can be expressed entirely in terms of the dimensionless couplings $\tilde{u}$, $w_u$, and $w_D$. To this order, we only have nontrivial $Z$-factors for $\Zutilde$ and $\Zwu$. Therefore, $w_D$ has no nontrivial flow function at this order, and the fixed points, along with the critical exponents, would depend on the bare value of $w_D$. This is an artifact of truncating the perturbative expansion at one-loop order. In what follows, we consider the two-loop diagrams renormalizing $D$, $w_D$ and $\Gamma$ in order to obtain a non-trivial flow function for $w_D$. Furthermore, including these will also allow us to compute the anomalous and dynamical exponents to lowest order in $\epsilon$. 

At two-loop order, only one diagram contributes to the renormalization of the noise vertex
\begin{equation} \label{eq:noisevertexnormalisation}
    \vertexnoise{}{} = 
    \begin{tikzpicture}[baseline = (o.base)]
  \begin{feynman}
    \vertex (o) at (0,0) ;
    \vertex [above left=0.6 of o] (i1) {\(a\)};
    \vertex [below left=0.6 of o] (i2) {\(b\)};
    \diagram* {
      (o) -- [scalar] (i1),
      (o) -- [plain] (i2)
    };
  \end{feynman}
\end{tikzpicture}
~+ 
\begin{tikzpicture}[baseline = (o.base)]
  \begin{feynman}
    \vertex (a) at (0,0) ;
    \vertex (b) at (0.8,0) ;
    \vertex (c) at (2,0);
    \vertex [above left=of a] (o1) {\(a\)};
    \vertex [below left=of a] (o2) {\(b\)};
    \vertex (a1) at ($(a)!0.7!(o1)$);
    \vertex (a2) at ($(a)!0.7!(o2)$);
    \diagram* {
      (o1) -- [scalar] (a1),
      (o2) -- [plain] (a2),
      (a1) -- [anti propagator] (a),
      (a1) -- [propagator] (b),
      (a1) -- [propagator] (c),
      (a2) -- [propagator] (a),
      (a2) -- [anti propagator] (b),
      (a2) -- [anti propagator] (c),
    };
  \end{feynman}
\end{tikzpicture}
\end{equation}
This diagram, denoted by $\diagram_{\noisediagram}^{ab}$, is given by
\begin{align} \label{eq:noise_integral}
    \diagram_{\noisediagram}^{ab} &= \kroneckerdelta{} \frac{\numcomponent+1}{9}u^2(1+w^2_u) \ddbarint{q_1} \ddbarint{q_2} \dbarint{\omega_1} \dbarint{\omega_2}C_0(\qvec_1, \omega_1)C_0(\qvec_2, \omega_2) C_0(\qvec_1 + \qvec_2, \omega_1 + \omega_2) = \kroneckerdelta{}\frac{n+1}{9}u^2(1+w_u)^2 \integral{\noisediagram} \nonumber \\
    &= 2\noise \kroneckerdelta{} \frac{\numcomponent+1}{9} \frac{u^2 \noise^2}{D^4}\left(1+\rationonlin^2\right) \Re\left[ \ddbarint{q_1} \ddbarint{q_2} \frac{1}{(\qvecsquared_1 + \tau/D)(\qvec_2^2 + \tau/D)((\qvec_1+\qvec_2)^2 + \tau/D)}\right. \times\nonumber \\
    & ~\left.\frac{1}{(1+\imag \ratiodiff)(\qvecsquared_1 +\qvecsquared_2) + (1-\imag \ratiodiff)(\qvec_1 + \qvec_2)^2 + 3\tau/D + \imag \tau'/D} \right].
\end{align}
Using several Feynman parameterizations, the wave-vector integrals can be evaluated. Specifically, we first use a Feynman parameterization to rewrite the wave-vector integral, $\integral{\noisediagram}$, in Eq.~\eqref{eq:noise_integral} as
\begin{equation} \label{eq:first_feynman_noise}
    \integral{\noisediagram} = \int_0^{1} \frac{dx}{(1+x)^2}\ddbarint{q_1} \frac{1}{\qvec^2_1 + \tau/\diff} \ddbarint{q_2} \frac{1}{(\qvec_2^2 + \tau/D)\left(\qvec^2_2 + 2\frac{1 -\imag x w_D}{1+x} \qvec_1 \cdot\qvec_2 + \qvec_1^2 + \frac{\tau}{\diff}\frac{1+2x+\imag \tau'/\tau x}{1+x}\right)^2}
\end{equation}
To evaluate the $\qvec_2$ integral, we introduce yet another Feynman parameterization, yielding
\begin{equation} \label{eq:second_feynman_noise}
    \integral{\noisediagram} = 2\int_0^{1} \frac{dx}{(1+x)^2} \int_0^{1}dy y\ddbarint{q_1} \frac{1}{\qvec^2_1 + \tau/\diff} \ddbarint{q_2} \frac{1}{\left(\qvec^2_2 + 2\frac{y(1 -\imag x w_D)}{1+x} \qvec_1 \cdot\qvec_2 + y\qvec_1^2 + \frac{\tau}{\diff}\left[\frac{y(1+2x+\imag \tau'/\tau x )}{1+x} +1 - y\right]\right)^3}
\end{equation}
Integral over $\qvec_2$ in Eq.~\eqref{eq:second_feynman_noise} can be evaluated using the identity \cite{tauber_critical_2014}:
\begin{equation} \label{Eq:integral_identity_2}
    \ddbarint k \frac{1}{(\kvecsquared + 2\qvec\cdot \kvec + \tau)^s} = \frac{\Gamma(s-d/2)}{\forpidtwo \Gamma(s)}\frac{1}{(\tau - \qvecsquared)^{s-d/2}}.
\end{equation}
Employing Eq.~\eqref{Eq:integral_identity_2} to evaluate the $\qvec_2$ integral in Eq.~\eqref{eq:second_feynman_noise} gives
\begin{equation}
    \integral{\noisediagram} = \frac{\Gamma(3-d/2)}{\forpidtwo}\int_0^{1} \frac{dx}{(1+x)^2} \int_0^{1}dy \frac{y}{\left[y\left( 1 - \frac{y(1 -\imag x w_D)^2}{(1+x)^2}\right) \right]^{3-d/2}}\ddbarint{q_1} \frac{1}{\qvec^2_1 + \tau/\diff} \frac{1}{\left(\qvec^2_1 + \frac{\tau}{\diff} \frac{\frac{y(1+2x+\imag \tau'/\tau x )}{1+x} +1 - y}{y\left( 1 - \frac{y(1 -\imag x w_D)^2}{(1+x)^2}  \right)}\right)^{3-d/2}}
\end{equation}
Employing a final Feynman parameterization to evaluate the $\qvec_1$ integral gives:
\begin{align} \label{eq:noisediagramfeynmanintegral}
    \diagram_{\noisediagram}^{ab} &= 2\noise \kroneckerdelta{} \frac{n+1}{9}\left( \frac{u \Gamma}{D^2 \forpidtwo}\right)^2 \left(1+w^2_u\right) \Gamma(4-d) \times \nonumber \\
    &\Re\left[\int_{0}^{1}\frac{dx}{(1+x)^2} \int_0^1 dy \int_0^{1} dz \frac{y z^{2-d/2}}{\left[y\left(1 - \frac{y(1-\imag x \ratiodiff)^2}{(1+x)^2}\right) \right]^{3-d/2}} \left(\frac{\tau}{D m(x,y,z)}\right)^{d-4} \right]
\end{align}
where 
\begin{equation}
    m(x,y,z) = z\frac{\frac{y(1+2x+\imag \tau'/\tau x )}{1+x} +1 - y}{y\left( 1 - \frac{y(1 -\imag x w_D)^2}{(1+x)^2}  \right)} + 1-z
\end{equation}
We are interested in the value of $\diagram_{\noisediagram}^{ab}$ to leading order in $1/\epsilon$ at the normalization point. Evaluating Eq.~\eqref{eq:noisediagramfeynmanintegral} at the NP and retaining only the leading divergence, tantamount to setting $d = 4$ inside the Feynman integrals, we find that the noise vertex at the NP is given by \cite{mathematica}:
\begin{equation} \label{eq:noisevertexfinal}
    \left. \vertexnoise{}{} \right|_{\rm NP} = 2\noise \kroneckerdelta{}\left\{ 1 + \frac{\numcomponent+1}{18}\frac{1+\rationonlin^2}{1+w^2_D}\left(\ln\left(\frac{4096}{(9+\ratiodiff^2)^3 (1+\ratiodiff^2)} \right)+2\ratiodiff\left( \arctan\left( \frac{\ratiodiff}{3}\right) + \arctan(\ratiodiff)\right)\right)\frac{\utilde^2 \mueps{-2}}{\epsilon} \right\} 
\end{equation}
Using Eqs.~\eqref{eq:renormconds}, \eqref{eq:renvertextonormalvertex} and \eqref{eq:noisevertexfinal}, we find that the $\Znoise$ satisfies the following relation
\begin{equation} \label{eq:Zfactornoise}
    \Znoise |\Zpsitilde|^2 = 1 + \frac{\numcomponent+1}{18}\frac{1+\rationonlin^2}{1+w^2_D}\left(\ln\left(\frac{4096}{(9+\ratiodiff^2)^3 (1+\ratiodiff^2)} \right)+2\ratiodiff\left( \arctan\left( \frac{\ratiodiff}{3}\right) + \arctan(\ratiodiff)\right)\right) \frac{\utilde^2 \mueps{-2}}{\epsilon}.
\end{equation}
As expected, we recover the Model A result upon setting $w_u = w_D = 0$. 

We now proceed with the renormalization of $D$ and $w_D$. In total, there are four two-loop diagrams contributing to $\invprop{}{}$; however, two of them renormalize only the mass and do not contribute to $Z_D$ or $Z_{D'}$. For our purposes, we only need to consider the renormalization of $D$ and $w_D$. Therefore, we only need to consider
\begin{equation}
    \diagram^{ab}_{\diffdiagramone}(\kvec,\omega) = 
\begin{tikzpicture}[baseline = (o.base), scale = 1.4]
\begin{feynman}
\vertex (o) at (0,0) {\(a\)};
\vertex (a1) at (0.4,0);
\vertex (a2) at ($(a1) + (1,0)$);
\vertex (a3) at ($(a2) + (0.33,1)$);     
\vertex (a4) at ($(a2) + (0.33,-1)$);    
\vertex (i) at ($(a2) + (1,0)$) {\(b\)};  
\diagram* {
(o) -- [scalar] (a1),
(a1) -- [propagator] (a2),
(a1) -- [propagator] (a3),
(a2) -- [anti propagator] (a3),
(a1) -- [anti propagator] (a4),
(a2) -- [propagator] (a4),
(a2) -- [plain, edge label= {\(\kvec,\omega\)}] (i),
};
\end{feynman}
\end{tikzpicture}
,~~~~\diagram_{\diffdiagramtwo}^{ab}(\kvec,\omega) = 
\begin{tikzpicture}[baseline = (o.base), scale = 1.4]
\begin{feynman}
\vertex (o) at (0,0) {\(a\)};
\vertex (a1) at (0.4,0);
\vertex (a2) at ($(a1) + (1,0)$);
\vertex (a3) at ($(a2) + (0.33,1)$);     
\vertex (a4) at ($(a2) + (0.33,-1)$);    
\vertex (i) at ($(a2) + (1,0)$) {\(b\)};  
\diagram* {
(o) -- [scalar] (a1),
(a1) -- [anti propagator] (a2),
(a1) -- [propagator] (a3),
(a2) -- [anti propagator] (a3),
(a1) -- [propagator] (a4),
(a2) -- [anti propagator] (a4),
(a2) -- [plain, edge label= {\(\kvec,\omega\)}] (i),
};
\end{feynman}
\end{tikzpicture}
\end{equation}
The first diagram is given by 
\begin{align}  \label{eq:inv_prop_one}
    \diagram_{\diffdiagramone}^{ab}(\kvec, \omega) &= \kroneckerdelta{} \frac{2(\numcomponent+1)}{9} \frac{u^2(1+\imag \rationonlin)^2 }{D^3} \ddbarint{q_1} \ddbarint{q_2} \dbarint{\omega_1} \dbarint{\omega_2} C_0(\qvec_1, \omega_1) C_0(\qvec_2, \omega_2) G_0(-\qvec_1 + \qvec_2 + \kvec, \omega + \omega_2 - \omega_1) \nonumber\\
    &= \kroneckerdelta{} \frac{2(\numcomponent+1)}{9} D(1+\imag \rationonlin)^2\frac{u^2\noise^2}{D^4} \ddbarint{q_1} \ddbarint{q_2} \frac{1}{(\qvecsquared_1 + \tau/D)(\qvecsquared_2 + \tau/D)} \times \nonumber \\
    & \frac{1}{-\imag \omega/D + (1+\imag \ratiodiff)(\qvecsquared_1 + (\qvec_1+\qvec_2+\kvec)^2) + (1-\imag \ratiodiff)\qvecsquared_2 + 3\tau/D + \imag \tau'/D}
\end{align}
Once again, through the use of several Feynman parameterizations, the wave-vector integrals can be evaluated. We start by collecting all the terms involving $\qvec_2$:
\begin{align}
    \integral{\diffdiagramone}(\kvec,\omega) &= \int_0^{1}\frac{dx}{(1+x)^2}\ddbarint{q_1} \frac{1}{\qvec_1^2 + \tau/D}\times \nonumber \\
    &\ddbarint{q_2} \frac{1}{\left(\qvec_2^2 + 2 \frac{x(1+\imag w_D)}{1+x}\qvec_2 \cdot (\kvec+\qvec_1) + \frac{x(1+\imag w_D)}{1+x}((\kvec+\qvec_1)^2 + \qvec_1^2) -\frac{\imag \omega}{\diff}\frac{x}{1+x} +\frac{\tau}{\diff}\frac{1+2x + \imag \tau'/\tau x}{1+x}\right)^2}
\end{align}
where once again we have defined $\integral{\diffdiagramone}(\kvec,\omega)$ as the wave-vector integrals we need to compute in Eq.~\eqref{eq:inv_prop_one}. Evaluating the integral using Eq.~\eqref{Eq:integral_identity_2}, expanding the denominator and collecting terms in terms of powers of $\qvec_2$ we obtain:
\begin{align}
    \integral{\diffdiagramone}(\kvec,\omega) &= \frac{\Gamma(2-d/2)}{\forpidtwo}\int_0^{1}\frac{dx}{(1+x)^2} \left[ \frac{(1+x)^2}{x(1+\imag \ratiodiff)(2+x(1-\imag \ratiodiff))}\right]^{2-d/2} \times \nonumber \\ 
    &\ddbarint{q_1} \frac{1}{\qvec_1^2 + \tau/D} \frac{1}{\left(\qvec_1^2 + 2 \frac{(1-\imag xw_D)}{2+x(1-\imag w_D)}\qvec_1 \cdot \kvec + \frac{(1-\imag xw_D)}{2+x(1-\imag w_D)}\kvec^2 -\frac{\imag \omega}{\diff} \frac{1+x}{(1+\imag w_D)(2+x(1-\imag w_D))}+\frac{\tau}{\diff}\frac{(1+x)(1+2x + \imag \tau'/\tau x)}{x(1 + \imag w_D)(2 + x(1-\imag w_D))}\right)^2}
\end{align}
Employing one final Feynman parameterization to bring the two factors into one single denominator and evaluating the resulting integral gives
\begin{align} \label{eq:diagramoneforinvprop}
    \diagram_{\diffdiagramone}^{ab}(\kvec, \omega) = \kroneckerdelta{} \frac{2(\numcomponent+1)}{9} D(1+\imag \rationonlin)^2 \utilde^2\Gamma(3-d) &\int_0^{1}\frac{dx}{(1+x)^2}\left[ \frac{(1+x)^2}{x(1+\imag \ratiodiff)(2+x(1-\imag \ratiodiff))}\right]^{2-d/2}\times \nonumber \\
    &\int_0^{1}dy y^{1-d/2}\frac{1}{\left[-\frac{\imag \omega}{D} \alpha_1(x,y) + \kvecsquared \beta_1(x,y) + \frac{\tau}{D}m_1(x,y)\right]^{3-d}}
\end{align}
where 
\begin{gather}
    \alpha_1(x,y) = \frac{y(1+x)}{(1+\imag \ratiodiff)(2+x(1-\imag \ratiodiff))}, ~\beta_1(x,y) = \frac{y(1-\imag x \ratiodiff)}{2+x(1-\imag \ratiodiff)}\left( 1 - \frac{y(1-\imag x \ratiodiff)}{2+x(1-\imag \ratiodiff)}\right), \nonumber \\
    ~m_1(x,y) = \frac{y(1+x)(1+2x + \imag\tau'/\tau x)}{x(1+\imag \ratiodiff)(2+x(1-\imag \ratiodiff))}.
\end{gather}
The second diagram, $\diagram_{\diffdiagramtwo}^{ab}(\kvec,\omega)$ is given by:
\begin{align} \label{eq:second_diagram_wd}
    \diagram_{\diffdiagramtwo}^{ab}(\kvec, \omega) &= \kroneckerdelta{} \frac{\numcomponent+1}{9} \frac{u^2(1+ \rationonlin^2) \noise^2}{D^3} \ddbarint{q_1} \ddbarint{q_2} \dbarint{\omega_1} \dbarint{\omega_2} C_0(\qvec_1, \omega_1) C_0(\qvec_2, \omega_2) \Gbar_0(\qvec_1 + \qvec_2 + \kvec, -\omega + \omega_2 + \omega_1) \nonumber\\
    &= \kroneckerdelta{} \frac{\numcomponent+1}{9} D(1+ \rationonlin^2)\frac{u^2\noise^2}{D^4} \ddbarint{q_1} \ddbarint{q_2} \frac{1}{(\qvecsquared_1 + \tau/D)(\qvecsquared_2 + \tau/D)} \times \nonumber \\
    & \frac{1}{-\imag \omega/D + (1+\imag \ratiodiff)(\qvecsquared_1 + \qvecsquared_2) + (1-\imag \ratiodiff)(\qvec_1+\qvec_2+\kvec)^2 + 3\tau/D + \imag \tau'/D}
\end{align}
Using similar Feynman parameterizations, the wave-vector integrals in Eq.~\eqref{eq:second_diagram_wd} can be evaluated to give:
\begin{align} \label{eq:diagramtwoforinvprop}
    \diagram_{\diffdiagramtwo}^{ab}(\kvec, \omega) = \kroneckerdelta{} \frac{\numcomponent+1}{9} D(1+ \rationonlin^2) \utilde^2\Gamma(3-d) &\int_0^{1}\frac{dx}{(1+x)^2}\left[ \frac{(1+x)^2}{x(1+\imag \ratiodiff)(2+x(1-\imag \ratiodiff))}\right]^{2-d/2}\times \nonumber \\
    &\int_0^{1}dy y^{1-d/2}\frac{1}{\left[-\frac{\imag \omega}{D} \alpha_2(x,y) + \kvecsquared \beta_2(x,y) + \frac{\tau}{D}m_2(x,y)\right]^{3-d}}
\end{align}
where 
\begin{gather} 
    \alpha_2(x,y) = \frac{y(1+x)}{2+x(1+2\imag \ratiodiff + \ratiodiff^2)}, ~\beta_2(x,y) = \frac{y(1-\imag \ratiodiff)(1+\imag x \ratiodiff)}{2+x(1+2\imag \ratiodiff + \ratiodiff^2)}\left( 1 - \frac{y(1-\imag \ratiodiff)(1+\imag x \ratiodiff)}{2+x(1+2\imag \ratiodiff + \ratiodiff^2)}\right), \nonumber \\
    ~m_2(x,y) = \frac{y(1+x)(1+2x + \imag\tau'/\tau x)}{x(2+x(1+2\imag \ratiodiff + \ratiodiff^2))} +1 -y.
\end{gather}
To obtain the correction to $-\imag \omega$ or $D(1+\imag \ratiodiff)\kvec^2$ in Eq.~\eqref{eq:treelevelvertex}, we take appropriate derivatives of $\diagram_{\diffdiagramone}^{ab}(\kvec, \omega)$ and $\diagram_{\diffdiagramtwo}^{ab}(\kvec, \omega)$
\begin{align} \label{eq:twoloopinversepropagator}
    \invprop{}{} = &\left\{-\imag \omega  \left(\kroneckerdelta{} - \left.\frac{\partial(\diagram^{ab}_{\diffdiagramone}(\kvec',\omega') +\diagram^{ab}_{\diffdiagramtwo}(\kvec',\omega'))}{\partial(-\imag \omega')} \right|_{\kvec' = 0,\omega' = 0}\right) \right. \nonumber \\
    & \left.+ D(1+\imag \ratiodiff)\kvecsquared\left(\kroneckerdelta{} - \left.\frac{\partial(\diagram^{ab}_{\diffdiagramone}(\kvec',\omega') +\diagram^{ab}_{\diffdiagramtwo}(\kvec',\omega'))}{\partial\kvec'^2} \right|_{\kvec' = 0,\omega' = 0}\right)\right\} + \invprop{}{\zerovec,0} + \mathcal{O}(\kvec^4, \omega \kvec^2, \omega^2).
\end{align}
Evaluating Eq.~\eqref{eq:twoloopinversepropagator} at the NP and using Eqs.~\eqref{eq:renormconds}-\eqref{eq:renvertextonormalvertex}, we obtain the following relations:
\begin{subequations}
    \begin{align}
        \Zpsi \Zpsitildecomplex &= 1 + \frac{\numcomponent+1}{9}(1+\imag \rationonlin)\frac{\utilde^2\mueps{-2}}{\epsilon}\int_{0}^{1}\frac{dx}{(1+x)^2}\int_{0}^{1}\frac{dy}{y}\left(2(1+\imag \rationonlin)\alpha_1(x,y) + (1-\imag \rationonlin) \alpha_2(x,y)\right) \\
        \Zpsi \Zpsitildecomplex \ZDiff(1+\imag \Zwu) &= (1+\imag \ratiodiff) \left(1 + \frac{\numcomponent+1}{9}\frac{1+\imag \rationonlin}{1+\imag \ratiodiff} \frac{\utilde^2 \mueps{-2}}{\epsilon} \int_{0}^{1}\frac{dx}{(1+x)^2}\int_{0}^{1}\frac{dy}{y}\left(2(1+\imag \rationonlin)\beta_1(x,y) + (1-\imag \rationonlin) \beta_2(x,y)\right)\right)
    \end{align}
\end{subequations}
Computing the integrals over $x$ and $y$ \cite{mathematica}, we obtain the following $Z-$factors:
\begin{subequations} \label{eq:Zfactorinvprop}
    \begin{align}
        &\Zpsi \Zpsitildecomplex = 1 + \frac{\numcomponent+1}{18}\frac{1+\imag \rationonlin}{(1+\ratiodiff^2)^2}\frac{\utilde^2\mueps{-2}}{\epsilon} \left( 4(\imag + \ratiodiff)^2(-\imag + \rationonlin) \arctan\left(\frac{\ratiodiff}{3}\right) - \log(9 + 10\ratiodiff^2 + \ratiodiff^4) \right. \nonumber \\
        & \left.- 4\imag(-\rationonlin + \ratiodiff(2 - 3\imag \ratiodiff + (6\imag + \ratiodiff)\rationonlin)) \log(2) + \log(4096) + 2(\imag + \ratiodiff)^2(1 + \imag \rationonlin) \log(9 + \ratiodiff^2) \right. \nonumber\\
        &\left. + 2(-\imag + \ratiodiff)^2(\imag + \rationonlin) \arctan\left(\frac{2\ratiodiff}{3 + \ratiodiff^2}\right)+ (\ratiodiff(-2\imag + \ratiodiff)(1 - \imag \rationonlin) + \imag \rationonlin) \log(9 + 10\ratiodiff^2 + \ratiodiff^4) \right), \\
        &\Zpsi \Zpsitildecomplex \ZDiff(1+\imag \Zwu) = (1+\imag \ratiodiff) \left( 1 + \frac{\numcomponent+1}{18}\frac{1+\imag \rationonlin}{1+\imag \ratiodiff} \frac{3+\imag(\rationonlin-2\ratiodiff)}{3-\imag \ratiodiff} \frac{\left(\utilde \mueps{-}\right)^2}{\epsilon}\right).
    \end{align}
\end{subequations}
Having determined all of the $Z$-factors to lowest non-trivial order, we proceed to identify the fixed points of the RG flow.

\subsection{Fixed Points of the RG flow}
The $Z-$factors in Eqs.~\eqref{eq:zfactormass}, \eqref{eq:zfactorsnonlincoupling}, \eqref{eq:Zfactornoise} and \eqref{eq:Zfactorinvprop} can be fully expressed in terms of $\utilde, \rationonlin$ and $\ratiodiff$. Therefore, these couplings control the fixed points of the theory. To identify the fixed points, we compute the $\beta$-functions governing the flow of ${\tilde{u}, w_u, w_D}$ with $\mu$:
\begin{subequations}
\begin{align} \label{eq:flowfuncdefn}
    \betautilde &\equiv \mu\frac{\partial \utilde_R}{\partial \mu} = \utildeR\left(-\epsilon + \mu\frac{\partial \ln\left( \Znonlincoupling\Znoise/\ZDiff^2\right)}{\partial \mu}\right), \\
    \betawu &\equiv \mu\frac{\partial w_{u,R}}{\partial \mu} = w_{u,R}\mu\frac{\partial \ln\left(\Zwu\right)}{\partial \mu}, \\
    \betawD &\equiv \mu\frac{\partial w_{D,R}}{\partial \mu} = w_{D,R}\mu\frac{\partial \ln\left(\ZwD\right)}{\partial \mu}
\end{align}
\end{subequations}
Using Eqs.~\eqref{eq:zfactorsnonlincoupling} and \eqref{eq:Zfactorinvprop}, we can calculate the flow functions to order $\utildeR^2$:
\begin{subequations} \label{eq:flowfunctions}
    \begin{align}
        \betautilde &= \utildeR\left( -\epsilon + \frac{2\utildeR}{3}\left( \numcomponent+4 - \frac{(\rationonlinR - \ratiodiffR)^2}{1+\ratiodiffR^2}\right)\right), \\
        \betawu &= \frac{2\utildeR}{3} \frac{\rationonlinR^2 + 1}{\ratiodiffR^2+1}(\rationonlinR - \ratiodiffR), \\
        \betawD &= \frac{2(\numcomponent+1)}{9}\frac{\utildeR^2}{(1+\ratiodiffR^2)(9 + \ratiodiffR^2)}\left[
    (1 + \ratiodiffR^2) (\ratiodiffR - \rationonlinR) (6 + \ratiodiffR (\ratiodiffR + \rationonlinR))  \right. \nonumber \\
    & + (9 + \ratiodiffR^2) (3 + 8 \ratiodiffR \rationonlinR - \rationonlinR^2 + \ratiodiffR^2 (-3 + \rationonlinR^2)) \arctan(\ratiodiffR/3)   \nonumber \\
    & + ( 9 + \ratiodiffR^2) ((-1 + \ratiodiffR^2) (1 + \rationonlinR^2) \arctan(\ratiodiffR) + 8 \rationonlinR \log(2) - 4 \ratiodiffR (1 + 2 \ratiodiffR \rationonlinR - 3 \rationonlinR^2) \log(2) \nonumber \\
    &  + \left. 2 (\ratiodiffR - \rationonlinR) (1 + \ratiodiffR \rationonlinR) \log(9+\ratiodiffR^2) - \ratiodiffR (1 + \rationonlinR^2) \log(9 + 10 \ratiodiffR^2 + \ratiodiffR^4))
    \right] . \label{eq:flowfunctionwd}
    \end{align}
\end{subequations}
where we have replaced bare coupling constants by their renormalized counterparts, valid to this order in the perturbation expansion. The fixed points, $\{\utildestar, \wDstar, \wustar\}$, correspond to the roots of the flow functions in Eqs.~\eqref{eq:flowfunctions}. We are interested in the stable fixed points of the model, where the stability matrix
\begin{equation} \label{eq:stabilitymat}
    \bm{\Lambda} = \begin{pmatrix}
        \frac{\partial \beta_{\utilde}}{\partial \utildeR} & \frac{\partial \beta_{\utilde}}{\partial w_{D,R}} & \frac{\partial \beta_{\utilde}}{\partial w_{u,R}} \\
        \frac{\partial \beta_{w_D}}{\partial \utildeR} & \frac{\partial \beta_{w_D}}{\partial w_{D,R}} & \frac{\partial \beta_{w_D}}{\partial w_{u,R}} \\
         \frac{\partial \beta_{w_u}}{\partial \utildeR} & \frac{\partial \beta_{w_u}}{\partial w_{D,R}} & \frac{\partial \beta_{w_u}}{\partial w_{u,R}} \\
    \end{pmatrix},
\end{equation}
has positive eigenvalues. 

As usual, the flow functions admit the Gaussian fixed point, $\{\utildestar = 0,\wDstar \in \Rset, \wustar \in \Rset\}$, which is stable for $d >4$. For $d<4$, another stable fixed point emerges. Specifically, solving $\betawu = \betautilde = 0$ gives
\begin{equation} \label{eq:crit_point_one_loop}
    \utildestar =\frac{3\epsilon}{2(\numcomponent+4)}, ~~ \wustar = \wDstar
\end{equation}
Finally, to find $\wDstar$, we solve $\betawD = 0$, with $\utildeR$ and $\rationonlinR$ evaluated at their fixed point values, Eq.~\eqref{eq:crit_point_one_loop}. This gives us the following equation for $\wDstar$:
\begin{align} \label{eq:fixedpointforwd}
    \frac{2(n+1)}{9}(\utildestar)^2&\left[ \wDstar\ln\left( \frac{16}{((\wDstar)^2 + 9)((\wDstar)^2+1)}\right) +  ((\wDstar)^2 -1)\arctan(\wDstar) \right.  \nonumber\\
    & +\left. ((\wDstar)^2 + 3)\arctan\left(\frac{\wDstar}{3}\right)  \right] = 0 
\end{align}
The function inside the square brackets is a monotonically increasing function of $\wDstar$, possessing a zero only at $\wDstar = 0$. Therefore, the nontrivial fixed point is given by 
\begin{equation} \label{eq:fixedpoint}
    \utildestar = \frac{3\epsilon}{2(\numcomponent+4)}, ~\wustar = \wDstar = 0 
\end{equation}
This fixed point corresponds to the Model~A fixed point. Evaluating the eigenvalues of the stability matrix in Eq.~\eqref{eq:stabilitymat}, one can show that it has no negative eigenvalues.

\subsection{Scaling Exponents}
At a fixed point of the RG flow, the system is scale invariant and physical observables exhibit power-law scaling. The scaling exponents of these observables define the universality class of the model and can be used to characterize the system. As usual, we are interested in the scaling behavior of the response and correlation functions. We show that they take the following form at the fixed point
\begin{subequations} \label{eq:scalingforms}
\begin{gather}
    \chi(\kvec,\omega) = \frac{1}{|\kvec|^{2-\eta}}\chihat\left(\frac{\omega}{|\kvec|^{z}}, a |\kvec|^{\eta-\eta_c},\tau |\kvec|^{-1/\nu} \ ,  \taup |\kvec|^{-1/\nu'}\right), ~ \\
    C(\kvec,\omega) = \frac{1}{|\kvec|^{2 + z-\eta'}}\Chat\left(\frac{\omega}{|\kvec|^{z}}, a|\kvec|^{\eta'-\eta_c},\tau |\kvec|^{-1/\nu}, \taup |\kvec|^{-1/\nu'}\right)
\end{gather}
\end{subequations}
where $a$ is some non-universal constant, $z$ is the dynamical scaling exponent quantifying the relative strength between time and space, $\nu$ and $
\nu'$ control the divergence of the correlation lengths and $\eta, \eta', \eta_c$ are anomalous exponents. Since detailed balance is broken at the microscopic level and there is no fluctuation-dissipation theorem, we allow for two independent anomalous exponents $\eta$ and $\eta'$. Finally, we have a novel exponent $\eta_c$ that controls the resonant frequency of the oscillations as a function of wave-vector \cite{tauber_perturbative_2014}.

We start the presentation of our results with the response function $\chi(\kvec,\omega)$, which is related to the vertex functions introduced previously as
\begin{equation}
    \chi(\kvec,\omega) = \left[ \frac{1}{n}\text{Tr}\{\bm{\Gamma}_{\psitildecomplex \psi}(\kvec,\omega)\}\right] ^{-1}.
\end{equation}
To derive its scaling form and the scaling exponents, we introduce explicitly the dependence of the response function on the renormalized parameters
\begin{equation}\elabel{responseR_from_response}
            Z^{-1}(\mu) 
\chi_{R}\left(\kvec,\omega;
            \tau_{R}(\mu),
            \tau'_R(\mu),
            D_R(\mu),  D'_R(\mu),
            \Gamma_R(\mu),
            \utildeR(\mu), w_{u,R}(\mu); 
            \mu\right) = \chi\left( \kvec,\omega,
            \tau,
            \tau',
            D, D',
            \Gamma,
            \utilde, w_{u}\right),
\end{equation}
where we have defined $Z = Z_{\psitildecomplex} Z_{\psi}$. Differentiating \Eref{responseR_from_response} with respect to $\mu$ leads us to the Callan-Symanzik equation \cite{bellac_quantum_1992}, which can be solved using the method of characteristics to yield
\begin{multline}\elabel{soln_CS_response}
            \ell^{-\zeta^{*}}
\chi_{R}\left(\kvec,\omega; 
            \ell^{\gamma^{*}_{\tau}}\tau_{R}(\mu_0),
            \ell^{\gamma^{*}_{\tau'}}\tau'_R(\mu_0),
            \ell^{\gamma^{*}_{D}}  D_R(\mu_0), \ell^{\gamma^{*}_{D'}}D'_R(\mu_0),
            \ell^{\gamma^{*}_{\noise}}\Gamma_R(\mu_0),
            \utildestar, \wustar; 
            \mu_0 \ell\right) \\
            = \chi_R\left(\kvec,\omega;
            \tau_{R}(\mu_0),
            \tau'_R(\mu_0),
              D_R(\mu_0), D'_R(\mu_0), \Gamma_R(\mu_0),
            \utildestar, \wustar; 
            \mu_0\right),
\end{multline}
close to the fixed point, where
\begin{subequations} \label{eq:defnwilson}
    \begin{alignat}{2}
       \zeta &= \mu \frac{\partial \log Z }{\partial \mu}  &\qquad \gamma_{\noise} &= \mu \frac{\partial \log Z_{\noise} }{\partial \mu}, \label{eq:wilsonfield} \\ 
        \gamma_{D} &= \mu \frac{\partial \log D_{R}}{\partial \mu} = \mu \frac{\partial \log Z_{D}}{\partial \mu}, &\qquad \gamma_{D'} &= \mu \frac{\partial \log D'_{R}}{\partial \mu} = \mu \frac{\partial \log Z_{D'}}{\partial \mu}\label{eq:wilsonD}\\
        \gamma_{\tau} &= \mu \frac{\partial \log \tau_{R}}{\partial \mu} = -2 + \mu \frac{\partial \log Z_{\tau}}{\partial \mu}, &\qquad \gamma_{\tau'} &= \mu \frac{\partial \log \tau'_{R}}{\partial \mu} = -2 + \mu \frac{\partial \log Z_{\tau'}}{\partial \mu} \label{eq:wilsontau}
    \end{alignat}
\end{subequations}
and the superscript $*$ signifies that the functions in \Erefs{defnwilson} are evaluated at the fixed point of the RG flow, \Eref{fixedpoint}. Furthermore, based on dimensional analysis, we have  
\begin{multline} \label{eq:dim_ana_res}
    \chi_{R}\left(\kvec,\omega;
            \tau_{R}(\mu),
            \tau'_R(\mu),
            D_R(\mu), D'_R(\mu),
            \Gamma_R(\mu),
            \utildeR, w_{u,R};
            \mu\right) = \\
            A^{-1} L^2 \chi_{R}\left( L \kvec, \frac{L^2 \omega}{A},
            \frac{\tau_{R}(\mu)}{A},
            \frac{\tau'_R(\mu)}{A},
            \frac{D_R(\mu)}{A}, \frac{D'_R(\mu)}{A},
            \frac{\Gamma_R(\mu)}{B},
            \utildeR, w_{u,R}; 
            \mu L\right) 
\end{multline}
for any $A,B$ and $L$. Using this on the LHS of \Eref{soln_CS_response} with length-scale $L = 1/(\mu_0\ell)$, noise scale $B = \noise_R(\mu_0) \ell^{\gamma^{*}_{\Gamma}}$ and diffusive scale $A = \ell^{\gamma^{*}_{D}}  D_R(\mu_0)$ gives 
\begin{equation} \label{eq:scalingform_response_derived}
    \chi_R(\kvec,\omega) = \frac{D^{-1}_R(\mu_0)}{\mu_0^2}\ell^{-2-\zetastar -\gammaDiffstar} \chi_R\left(\frac{\kvec}{\mu_0\ell},\frac{\omega}{D_R(\mu_0)\mu_0^2 \ell^{2 + \gammaDiffstar}}\frac{\tau_{R}(\mu_0) \ell^{\gammataustar - \gammaDiffstar}}{D_{R}(\mu_0)}, \frac{\tau'_{R}(\mu_0) \ell^{\gammataupstar - \gammaDiffstar}}{D_{R}(\mu_0)},1, a |\kvec|^{\gammaDiffpstar - \gammaDiffstar},1, \utildestar, \wustar ;1\right)
\end{equation}
where we have defined $a = D'_{R}(\mu_0)/D_{R}(\mu_0)$. Further choosing $\ell = |\kvec|/\mu_0$ yields the scaling form in \Eref{scalingforms} and quoted in the main text. Comparing \Erefs{scalingforms} and \eref{scalingform_response_derived} allows us to identify the scaling exponents in terms of the Wilson functions evaluated at the stable fixed point:
\begin{subequations} \label{eq:first_set_scaling}
\begin{align}
    z &= 2 + \gammaDiffstar, \\
    1/\nu &= \gammaDiffstar - \gammataustar, \\
    1/\nu' &= \gammaDiffstar - \gammataupstar, \\
    \eta &= -\zetastar - \gammaDiffstar, 
    \\
    \eta_c &= -\zetastar - \gammaDiffpstar.
\end{align} 
\end{subequations}

An analogous analysis can be done for the correlation function to determine $\eta'$. Following the same approach as in \Eref{responseR_from_response}, we introduce the renormalized correlation function with the full set of renormalized parameters
\begin{equation}\elabel{CorrR_from_Corr}
            |Z_{\psivec}(\mu)|^{-2}
C_{R}(
            \kvec,\omega; \tau_{R}(\mu),
            \tau'_{R}(\mu),
            D_{R}(\mu),
            D'_{R}(\mu),
            \Gamma_{R}(\mu),
            \utilde_{R}(\mu),
            w_{u,R}(\mu); 
            \mu) = C(\kvec,\omega;
             \tau,
             \tau',
             D,
             D',
             \Gamma,
             u,
             u'),
\end{equation}
Differentiating the LHS of \Eref{CorrR_from_Corr} with respect to $\mu$ leads us to another Callan-Symanzik equation, now for the correlation function, which can also be solved by the method of characteristics as before, leading to
\begin{multline}\elabel{CSeqn_soln_corr}
                \ell^{-\gammapsicomplexstar - \gammapsistar }
C_{R}\big(\kvec,\omega;
            \ell^{\gammataustar} \tau_{R}(\mu_0),
            \ell^{\gammataupstar} \tau'_{R}(\mu_0),
            \ell^{\gamma^{*}_{D}} D_{R}(\mu_0),
            \ell^{\gamma^{*}_{D'}} D'_{R}(\mu_0), 
            \ell^{\gamma^{*}_{\Gamma}} \Gamma_{R}(\mu_0),
            \utildestar,
            \wustar,
            \{\kvec\},\{\omega\}; 
            \ell \mu_0 \big)\\         
=
C_{R}\big(\kvec,\omega;
            \tau'_{R}(\mu_0),
            \tau'_{R}(\mu_0),
            D_{R}(\mu_0),
            D'_{R}(\mu_0),
            \Gamma_{R}(\mu_0),
            \utildestar,
            \wustar;
            \mu_0)        
\end{multline}
close to the critical point, where the coupling constants have reached their fixed point values. Furthermore, based on dimensional analysis, we have
\begin{multline} \label{eq:dim_ana_corr}
    C_{R}(\kvec,\omega;
            \tau_{R},
            \tau'_{R},
            D_R,
            D'_{R},
            \Gamma_{R},
            \utildestar,
            \wustar,
            \{\kvec\},\{\omega\}; 
            \mu)
\\ =
 B A^{-2} L^{4} 
    C_{R}\bigg(L \kvec, \frac{L^2}{A}\omega;
            \frac{\tau_{R}}{A},
            \frac{\tau'_{R}}{A},
            \frac{D_R}{A},
            \frac{D'_{R}}{A},
            \frac{\Gamma_{R}}{B},
            \utildestar,
            \wustar; 
            \mu L\bigg)
\end{multline}
for any $B, L$ and $A$. Once again, using \Eref{dim_ana_corr} on the LHS of \Eref{CSeqn_soln_corr} with length-scale $L = 1/(\ell \mu_0)$, diffusive scale $A = D_{R}(\mu_0) \ell^{\gammaDiffstar}$ and noise scale $B = \noise_R(\mu_0) \ell^{\gammanoisestar}$ gives the scaling form 
\begin{multline}
    C_R(\kvec,\omega) = \nonumber \\
    \frac{\noise_R(\mu_0)}{ \mu_0^4 D^2_{R}(\mu)}\ell^{-4 -\gammapsicomplexstar - \gammapsistar + \gammanoisestar - 2\gammaDiffstar}\hat{C}_R\left(\frac{|\kvec|}{\ell \mu_0}, \frac{\omega}{ \mu_0^2 D_{R}(\mu_0)\ell^{2+\gammaDiffstar}}, \frac{\tau_{R}(\mu_0)}{D_{R}(\mu_0)}\ell^{\gammataustar - \gammaDiffstar},  \frac{\tau'_{R}(\mu_0)}{D_{R}(\mu_0)}\ell^{\gammataupstar - \gammaDiffstar},1, \frac{D'_{R}(\mu_0)}{D_{R}(\mu_0)}\ell^{\gammaDiffpstar - \gammaDiffstar}, 1, \utildestar, \wustar; 1\right).
\end{multline}
Further choosing $\ell = |\kvec|/\mu_0$ gives us the advertised scaling form in Eq.~\eqref{eq:scalingforms}, allowing us to identify
\begin{equation}
    4 +\gammapsicomplexstar + \gammapsistar - \gammanoisestar + 2\gammaDiffstar = 2 + z - \eta'.
\end{equation}
Using the scaling exponents in \Eref{first_set_scaling}, we can find the following expression for $\eta'$:
\begin{align} \label{eq:exponentetap}
    \eta'&= -\gammapsicomplexstar-\gammapsistar +  \gammanoisestar - \gammaDiffstar \nonumber \\
    & =\left(\gammanoisestar + \gammapsitildestar + \gammapsitildecomplexstar \right) - 2\Re[\zetastar] - \gammaDiffstar
\end{align}
where we have added and subtracted $\gammapsitildecomplexstar$ and $\gammapsitildestar$ in going from the first line to the second line. Using the $Z-$ factors in Eqs.~\eqref{eq:zfactormass},  \eqref{eq:Zfactorinvprop} and \eqref{eq:Zfactornoise}, we can calculate the scaling exponent to lowest nontrivial order in $\epsilon$ at the stable fixed point in Eq.~\eqref{eq:fixedpoint}:
\begin{subequations}
\begin{align}
    z &= 2 + \frac{n+1}{4(n+4)^2}\left(6 \ln\left(\frac{4}{3}\right) - 1 \right) \epsilon^2 + \mathcal{O}(\epsilon^3),\\
    \eta &= \frac{n+1}{4(n+4)^2}\epsilon^2 + \mathcal{O}(\epsilon^3),\\
    \eta'&=\frac{n+1}{4(n+4)^2}\epsilon^2 + \mathcal{O}(\epsilon^3), \\
    \eta_c &= -\frac{n+1}{4(n+4)^2}\left(4 \ln\left(\frac{4}{3}\right) - 1 \right)\epsilon^2 + \mathcal{O}(\epsilon^3),~ \\
    \nu &= \frac{1}{2} + \frac{n+1}{4(n+4)}\epsilon + \mathcal{O}(\epsilon^2), \\
    \nu' &= \frac{1}{2} + \mathcal{O}(\epsilon^2)
\end{align}
\end{subequations}
Remarkably, apart from the novel exponent $\eta_c$, we recover the scaling exponents of Model A with $2\numcomponent$ field components. Specifically, at the fixed point we have $\gammanoisestar + \gammapsitildecomplexstar + \gammapsitildestar = \zetastar$, ensuring $\eta = \eta'$. Even though the microscopic dynamics are inherently non-reciprocal and break detailed balance, the asymptotic dynamics are agnostic to this and the transition to the oscillatory phase is governed by the Model~A universality class.

\section{Large Number of Vector Components} \label{sec:LargeN}
In this section, we analyze the Langevin Eqs.~\eqref{eq:modelv2supp} in the limit $\numcomponent \to \infty$. In the traditional $O(\numcomponent)$ models, this limit maps the theory onto the spherical model \cite{Spherical_model_stanley, spherical_model_kac}, where exact results can be obtained \cite{mosheQuantumFieldTheory2003a, zinn-justin_quantum_2002}. 

To analyze the limit, we first rescale $u$ and $u'$ by $\numcomponent$ in Eq.~\eqref{eq:modelv2supp}, allowing for a well-defined limit
\begin{equation} \label{eq:model_again}
     \dot{\psivec}(\xvec,t) = - \left\{ r+\imag r' -(\diff + \imag\diffp)\nablasquared + \frac{u + \imag u'}{n} \psiveccomplex(\xvec,t)\cdot  
        \psivec(\xvec,t)\right\} \psivec(\xvec,t) + \sqrt{2\Gamma}\xivec(\xvec,t) \ ,
\end{equation}
where we have absorbed the numerical factor $3$ into the nonlinear couplings $u$ and $u'$ for convenience. In the limit $\numcomponent \to \infty$, by the law of large numbers, we can replace $\psiveccomplex(\xvec,t)\cdot \psivec(\xvec,t)/n$ by its average:
\begin{equation}
    \frac{1}{\numcomponent}\psiveccomplex(\xvec,t)\cdot \psivec(\xvec,t) \overset{\numcomponent\to\infty }{=} \frac{1}{n}\left\langle \psiveccomplex(\xvec,t)\cdot \psivec(\xvec,t)\right\rangle + \mathcal{O}\left(\frac{1}{\sqrt{\numcomponent}}\right)
\end{equation}
whence the dynamics in Eq.~\eqref{eq:model_again} becomes
\begin{equation} \label{eq:large_n_limit}
    \dot{\psivec}(\xvec,t) = - \left\{ r+\imag r' -(\diff + \imag\diffp)\nablasquared + (u + \imag u') C(\zerovec, 0)\right\} \psivec(\xvec,t) + \sqrt{2\Gamma}\xivec(\xvec,t) \ .
\end{equation}
where we have used the definition of the correlation function to replace $\frac{1}{n}\left\langle \psiveccomplex(\xvec,t)\cdot \psivec(\xvec,t)\right\rangle$.  Since the equation of motion is linear, the correlation and response functions associated with Eq.~\eqref{eq:large_n_limit} can be found
\begin{subequations} \label{eq:large_n_corr_res}
\begin{align}
    \chi^{-1}(\kvec,\omega) &= -\imag \omega +(\diff + \imag \diffp)\kvecsquared + r+\imag r' + (u+\imag u') g(r,r')\label{eq:large_n_res} \\
    C(\kvec,\omega) &=\frac{2\noise}{(\omega - r'-\diffp\kvecsquared + u'g(r,r'))^2 +(\diff\kvecsquared + r + ug(r,r'))^2} \label{eq:large_n_corr}
\end{align} 
\end{subequations}
where we have defined the real quantity $g(r,r') \equiv C(\zerovec,0)$, whose self-consistent equation is shown below. Since $g(r,r')$ is independent of $\kvec$ and $\omega$, we can infer the values of $z, \eta, \eta_c$ and $\eta$ immediately from Eqs.~\eqref{eq:large_n_corr_res} by massaging it to the scaling form in Eqs.~\eqref{eq:scalingforms}:
\begin{equation}
    z = 2, ~~ \eta = \eta'= \eta_c = 0.
\end{equation}
To compute the exponents $\nu$ and $\nu'$, we have to evaluate Eq.~\eqref{eq:large_n_res} for small $\tau \equiv r-r_c$ and $\tau'= r'-r'_c$. We can identify the critical point $r_c, r'_c$ from Eqs.~\eqref{eq:criticalpoint}:
\begin{equation}  \label{eq:critical_point_large_n}
    \left. \chi^{-1}(\kvec = \zerovec,\omega=0) \right|_{r = r_c, r'= r'_c} = 0 \longrightarrow r_c + \imag r'_c + (u+\imag u')g(r_c,r'_c) = 0
\end{equation}
The function $g(r,r')$ is defined self-consistently using its definition and Eq.~\eqref{eq:large_n_corr_res}
\begin{equation} \label{eq:self_consistency_g}
    g(r,r') = C(\zerovec,0) = \ddbarint{k}\dbarint{\omega}C(\kvec,\omega) = \ddbarint{k}\frac{\noise}{D\kvecsquared + r + ug(r,r')}
\end{equation}
Eq.~\eqref{eq:critical_point_large_n} can be separated into a real and an imaginary part:
\begin{subequations} \label{eq:critical_point_large_n_re_im}
    \begin{align}
        r_c + ug(r_c,r'_c) = 0 \\
        r'_c + u'g(r_c,r'_c) = 0 
    \end{align}
\end{subequations}
After having identified the critical point, we can start parameterizing Eq.~\eqref{eq:large_n_res} in terms of $\tau$ and $\tau'$
\begin{align} \label{eq:large_n_res_tau}
    \chi^{-1}(\kvec,\omega) &= -\imag \omega +(\diff + \imag \diffp)\kvecsquared + \tau + \imag \tau' + (u+\imag u') (g(r_c + \tau,r_c' + \tau') - g(r_c,r'_c)) \nonumber \\
    &= -\imag \omega + (\diff + \imag \diffp)\kvecsquared + f(\tau,\tau')  
\end{align}
where we have used Eqs.~\eqref{eq:critical_point_large_n_re_im} to eliminate $r_c$ and $r'_c$ and defined 
\begin{equation}
    f(\tau,\tau') = \tau + \imag \tau' + (u+\imag u') (g(r_c + \tau,r_c' + \tau') - g(r_c,r'_c))
\end{equation}
To find expressions for the exponents $\nu$ and $\nu'$, we need to analyze the behavior of $f(\tau,\tau')$ for small $\tau$ and $\taup$. Using Eq.~\eqref{eq:self_consistency_g}, we can derive a self-consistency equation for the real and imaginary part of $f(\tau,\tau') = f_R(\tau,\tau') + \imag f_I(\tau,\tau')$
\begin{subequations} \label{eq:self_consistency_f}
\begin{align} 
    f_R(\tau,\tau') &= \tau + \frac{u\noise}{D}\ddbarint{k}\left[\frac{1}{ \kvecsquared + f_R(\tau,\tau')/D} - \frac{1}{\kvecsquared}\right] \ , \label{eq:self_consistency_f_re} \\
    f_I(\tau,\tau') &= \taup + \frac{u'\noise}{D}\ddbarint{k}\left[\frac{1}{ \kvecsquared + f_R(\tau,\tau')/D} - \frac{1}{\kvecsquared}\right] \ .\label{eq:self_consistency_f_im}
\end{align}
\end{subequations}
Since no explicit $\tau'$ appears in the self-consistency equation for $f_R(\tau,\tau')$, it has no dependence on $\tau'$ and is only a function of $\tau$; we therefore write $f_R(\tau,\tau') \equiv f_R(\tau)$. Consequently, by Eq.~\eqref{eq:self_consistency_f_im}, $f_{I}(\tau,\tau')$ is a linear function of $\tau'$. Evaluating the wave-vector integral in Eq.~\eqref{eq:self_consistency_f_re} using Eq.~\eqref{eq:integral_identity}, we obtain
\begin{equation} \label{eq:self_consistency_f_re_2}
    f_R(\tau) = \tau + \frac{u\noise}{D}\frac{\Gamma(1-d/2)}{\forpidtwo}f_R^{d/2-1}(\tau)
\end{equation}
We are only interested in the leading-order contribution of $\tau$ to $f_R(\tau)$. Assuming $f_R(\tau) \sim \tau^{\alpha}$ for small $\tau$, and using Eq.~\eqref{eq:self_consistency_f_re_2}, we can show that for $2<d<4$:
\begin{equation}\label{eq:soln_to_re}
    f_R(\tau) = \left(-\frac{\forpidtwo D}{\Gamma(1-d/2)u\noise} \right)^{2/(d-2)} \tau^{2/(d-2)} + \dots
\end{equation}
where the dots include higher-order terms in $\tau$. By using Eqs.~\eqref{eq:self_consistency_f_re} and \eqref{eq:soln_to_re}, we can find also a solution for $f_I(\tau,\tau')$, valid for small $\tau$
\begin{equation} \label{eq:soln_to_im}
    f_I(\tau,\tau') = \tau' + \frac{u'}{u}\left(-\frac{\forpidtwo D}{\Gamma(1-d/2)u\noise} \right)^{2/(d-2)} \tau^{2/(d-2)} + \dots
\end{equation}
where once again the dots include higher order terms in $\tau$ and $\tau'$. Using Eqs.~\eqref{eq:soln_to_re} and \eqref{eq:soln_to_im}, we can show that close to criticality, the response function is given by:
\begin{equation} \label{eq:res_close_to_criticality}
    \chi^{-1}(\kvec,\omega) = -\imag \omega +(\diff + \imag \diffp)\kvecsquared + b\tau^{2/(d-2)} + \imag \tau'
\end{equation}
where $b$ is some number. Massaging Eq.~\eqref{eq:res_close_to_criticality} into the scaling form in Eq.~\eqref{eq:scalingforms}, we can identify the scaling exponents $\nu$ and $\nu'$:
\begin{equation}
    \nu = \frac{1}{d-2}, ~~~ \nu'= \frac{1}{2}.
\end{equation}
We find that the scaling exponent $\nu$ is given by the equilibrium value \cite{zinn-justin_quantum_2002, mosheQuantumFieldTheory2003a} and $\nu'$ retains its mean-field value even in this limit.

\bibliography{NRwithU_n_symmetry}